\documentclass[twocolumn]{aastex631}

\usepackage{CJK}

\newcommand{\thisstar}{KELT-20}
\newcommand{\thisplanet}{KELT-20~b}

\shorttitle{Search for Thermal Inversion Agents in KELT-20 b}
\shortauthors{Johnson et al.}
\graphicspath{{./}{figures/}}

\begin{document}
\begin{CJK*}{UTF8}{gbsn}

\title{The PEPSI Exoplanet Transit Survey (PETS).\footnote{Based on data acquired with the Potsdam Echelle Polarimetric and Spectroscopic Instrument (PEPSI)
using the Large Binocular Telescope (LBT) in Arizona.} 
\\
II. A Deep Search for Thermal Inversion Agents in KELT-20 b/MASCARA-2 b with Emission and Transmission Spectroscopy}

\correspondingauthor{Marshall C. Johnson}
\email{johnson.7240@osu.edu}

\author[0000-0002-5099-8185]{Marshall C. Johnson} 
\affiliation{Department of Astronomy, The Ohio State University, 4055 McPherson Laboratory, 140 West 18$^{\mathrm{th}}$ Ave., Columbus, OH 43210 USA}

\author[0000-0002-4361-8885]{Ji Wang (王吉)}
\affiliation{Department of Astronomy, The Ohio State University, 4055 McPherson Laboratory, 140 West 18$^{\mathrm{th}}$ Ave., Columbus, OH 43210 USA}

\author[0000-0002-8823-8237]{Anusha Pai Asnodkar} 
\affiliation{Department of Astronomy, The Ohio State University, 4055 McPherson Laboratory, 140 West 18$^{\mathrm{th}}$ Ave., Columbus, OH 43210 USA}

\author{Aldo S. Bonomo}
\affiliation{INAF -- Osservatorio Astrofisico di Torino, Strada Osservatorio 20, I-10025 Pino Torinese, Italy}

\author[0000-0003-0395-9869]{B. Scott Gaudi}
\affiliation{Department of Astronomy, The Ohio State University, 4055 McPherson Laboratory, 140 West 18$^{\mathrm{th}}$ Ave., Columbus, OH 43210 USA}

\author[0000-0002-1493-300X]{Thomas Henning}
\affiliation{Max-Planck-Institut f\"ur Astronomie, K\"onigstuhl 17, D-69117 Heidelberg, Germany}

\author{Ilya Ilyin}
\affiliation{Leibniz-Institute for Astrophysics Potsdam (AIP), An der Sternwarte 16, D-14482 Potsdam, Germany}

\author{Engin Keles}
\affiliation{Leibniz-Institute for Astrophysics Potsdam (AIP), An der Sternwarte 16, D-14482 Potsdam, Germany}

\author{Luca Malavolta}
\affiliation{INAF -- Osservatorio Astronomico di Padova, Vicolo dell'Osservatorio 5, I-35122 Padova, Italy}
\affiliation{Dipartimento di Fisica e Astronomia ``Galileo Galilei'', Universita degli Studi di Padova, I-35122 Padova, Italy }

\author[0000-0003-2865-042X]{Matthias Mallonn}
\affiliation{Leibniz-Institute for Astrophysics Potsdam (AIP), An der Sternwarte 16, D-14482 Potsdam, Germany}

\author[0000-0002-0502-0428]{Karan Molaverdikhani}
\affiliation{Universit\"ats-Sternwarte, Ludwig-Maximilians-Universit\"at M\"unchen, Scheinerstrasse 1, D-81679 M\"unchen, Germany}
\affiliation{Exzellenzcluster Origins, Boltzmannstra{\ss}e 2, 85748 Garching, Germany}
\affiliation{Max-Planck-Institut f\"ur Astronomie, K\"onigstuhl 17, D-69117 Heidelberg, Germany}

\author{Valerio Nascimbeni}
\affiliation{INAF -- Osservatorio Astronomico di Padova, Vicolo dell'Osservatorio 5, I-35122 Padova, Italy}

\author{Jennifer Patience}
\affiliation{School of Earth and Space Exploration, Arizona State University, 660 S. Mill Ave., Tempe, Arizona 85281, USA}

\author{Katja Poppenhaeger}
\affiliation{Leibniz-Institute for Astrophysics Potsdam (AIP), An der Sternwarte 16, D-14482 Potsdam, Germany}
\affiliation{Institute of Physics \& Astronomy, University of Potsdam, Karl-Liebknecht-Str. 24/25, D-14476 Potsdam, Germany}

\author[0000-0003-2029-0626]{Gaetano Scandariato}
\affiliation{INAF -- Osservatorio Astrofisico di Catania, via S. Sofia 78, I-95123 Catania, Italy}

\author{Everett Schlawin}
\affiliation{Steward Observatory, University of Arizona, 933 N. Cherry Ave., Tucson, AZ 85721, USA}

\author{Evgenya Shkolnik}
\affiliation{School of Earth and Space Exploration, Arizona State University, 660 S. Mill Ave., Tempe, Arizona 85281, USA}

\author{Daniela Sicilia}
\affiliation{INAF -- Osservatorio Astrofisico di Catania, via S. Sofia 78, I-95123 Catania, Italy}

\author{Alessandro Sozzetti}
\affiliation{INAF -- Osservatorio Astrofisico di Torino, Strada Osservatorio 20, I-10025 Pino Torinese, Italy}

\author{Klaus G. Strassmeier}
\affiliation{Leibniz-Institute for Astrophysics Potsdam (AIP), An der Sternwarte 16, D-14482 Potsdam, Germany}
\affiliation{Institute of Physics \& Astronomy, University of Potsdam, Karl-Liebknecht-Str. 24/25, D-14476 Potsdam, Germany}

\author{Christian Veillet}
\affiliation{Large Binocular Telescope Observatory, 933 N. Cherry Ave., Tucson, AZ 85721, USA}

\author[0000-0001-9585-9034]{Fei Yan}
\affiliation{Institute of Astrophysics, University of G\"ottingen, Friedrich-Hund-Platz 1 , D-37077 G\"ottingen, Germany}
\affiliation{Max-Planck-Institut f\"ur Astronomie, K\"onigstuhl 17, D-69117 Heidelberg, Germany}



\begin{abstract}

Recent observations have shown that the atmospheres of ultra hot Jupiters (UHJs) commonly possess temperature inversions, where the temperature increases with increasing altitude. Nonetheless, which opacity sources are responsible for the presence of these inversions remains largely observationally unconstrained.
We used LBT/PEPSI to observe the atmosphere of the UHJ \thisplanet\ in both transmission and emission in order to search for molecular agents which could be responsible for the temperature inversion. We validate our methodology by confirming previous detections of Fe~\textsc{i} in emission at $16.9\sigma$.
Our search for the inversion agents TiO, VO, FeH, and CaH results in non-detections. Using injection-recovery testing we set $4\sigma$ upper limits upon the volume mixing ratios for these constituents as low as $\sim1\times10^{-9}$ for TiO.
For TiO, VO, and CaH, our limits are much lower than expectations from an equilibrium chemical model, while we cannot set constraining limits on FeH with our data. We thus rule out TiO and CaH as the source of the temperature inversion in \thisplanet, and VO only if the line lists are sufficiently accurate.

\end{abstract}



\section{Introduction} \label{sec:intro}

Hot Jupiters (HJs) offer some of our best chances at present to characterize exoplanetary atmospheres in detail. These planets' large radii, hot temperatures, short orbital periods, and often bright host stars mean that we can obtain more frequent observations and overall higher signal-to-noise ratios than are possible for smaller, cooler, or longer-period planets.

Several ultra hot Jupiters (UHJs; planets with equilibrium temperatures in excess of 2200 K) have recently been shown to have atmospheric temperature inversions through the presence of emission lines of atomic species in their spectra; if there was no inversion, the spectrum would present absorption lines instead. A temperature inversion occurs when some absorber exists within the atmosphere, heating the atmosphere and causing the temperature to increase, rather than decrease, with increasing altitude. In order to cause an inversion, an absorber needs to have a very large optical or ultraviolet cross section in order to deposit a significant amount of energy within a relatively thin layer of the atmosphere. UHJs had previously been proposed to harbor thermal inversions, with the transition temperature in this regime occurring anywhere from 1700 to 2500 K \citep[e.g.,][]{Baxter20,Lothringer18}. 
Individual planets with detections of line emission due to an inversion include WASP-33~b \citep[e.g.,][]{Nugroho17}, WASP-121~b \citep{Evans17}, WASP-18~b \citep{Sheppard17}, WASP-189~b \citep[e.g.,][]{Yan20,Yan22-W33W189}, KELT-9~b \citep[e.g.,][]{Pino20}, and \thisplanet\ \citep{Cont21,Fu22,Borsa22,Kasper22}. The number of possible inversion agents in these planets is limited, as only a few molecular species can survive at these high temperatures.

Nonetheless, direct evidence for which species cause these inversions has been lacking. \cite{Nugroho17}, \cite{Nugroho20}, and \cite{Cont21-W33} detected TiO in WASP-33~b, but \cite{Herman20} and \cite{Serindag21} were unable to confidently detect TiO with their own data or methods. \cite{Prinoth22} detected TiO in absorption in WASP-189~b. VO has been detected at low resolution in WASP-121~b, but high-resolution observations have been unable to confirm this detection, likely due to inaccuracies with the VO line list \citep{Hoeijmakers20-W121,Merritt21,deRegt22}. Additionally, however, low-resolution observations generally probe lower pressure levels than high-resolution data, which could have different chemistry. No molecular inversion agent candidate has been definitively detected in either KELT-9~b or \thisplanet\ despite the presence of atomic emission lines, motivating further searches. KELT-9~b, however, may be too hot for any molecular agent to exist, but \cite{ChangeatEdwards21} found evidence for molecular absorption at low resolution. They attributed this absorption to TiO, VO, FeH, and H$^-$, but none of these species have been individually confirmed at high resolution. Cooler planets have also been subject to extensive searches for TiO, most notably WASP-76~b; \cite{Tsiaras18} found evidence for TiO and VO, and \cite{vonEssen20} marginal evidence for TiH. \cite{Edwards20}, however, showed that these were due to an unaccounted-for stellar companion but still found evidence for an atmospheric inversion, and \cite{Tabernero21} found no evidence of these species at high resolution. 

A number of molecular species, including TiO, VO \citep{Fortney08}, SiO, metal hydrides \citep{Lothringer18}, AlO, CaO, NaH, and MgH \cite{GandhiMadhusudhan19}, have been proposed to cause inversions. Alternately, in the hottest HJs, atomic species such as Fe or Mg, or continuum H$^-$ opacity, could give rise to inversions \citep{Lothringer18}. All of these species fulfill the criterion of possessing large optical or UV cross-sections that could give rise to inversions.

KELT-20~b/MASCARA-2~b (hereafter \thisplanet\ for brevity) is an UHJ discovered independently by both the Kilodegree Extremely Little Telescope (KELT) and Multi-site All-Sky
CAmeRA (MASCARA) ground-based transit surveys \citep{Lund17,Talens18}. With a $V$-band magnitude of 7.58, the host star HD 185603 is one of the five brightest stars to host a known transiting giant planet. It has thus been a target for atmospheric observations using using broadband phase curves, high-resolution transmission spectroscopy, and emission spectroscopy at both high and low resolution \citep[e.g.,][]{Wong21,CasasayasBarris18,Cont21,Fu22}. 

Many observations of the atmosphere of \thisplanet\ have already been obtained via high-resolution transmission spectroscopy. These have allowed the detection of several atomic species in the planet's atmosphere: H \textsc{i}, Na \textsc{i}, \citep{CasasayasBarris18}, Ca \textsc{ii}, Fe \textsc{ii} \citep{CasasayasBarris19,Nugroho20,BelloArufe22}, Fe \textsc{i} \citep{Stangret20}, Mg \textsc{i}, and Cr \textsc{ii} \citep{Hoeijmakers20}. Critically, however, \cite{Nugroho20} reported non-detections of the potential thermal inversion agents NaH, MgH, AlO, SH, CaO, VO, FeH, and TiO, while \cite{Kesseli20} found a low-confidence signal for FeH. 

An alternative method is to use emission spectroscopy. Although the emission signals are overall smaller than the transmission signals, 
emission spectroscopy directly probes the altitudes around the temperature inversion, since this is where the emission lines must be formed.

Several species have already been detected in emission for \thisplanet\ at high resolution: Fe \textsc{i} \citep{Yan22,Borsa22}, Si \textsc{i} \citep{Cont21}, Fe \textsc{ii}, Cr \textsc{i} \citep{Borsa22}, and \textsc{Ni i} \citep{Kasper22}.
\cite{Fu22} also detected H$_2$O and CO emission at low resolution. These detections of emission lines not only add to the detected atmospheric species but also demonstrate the presence of an inverted temperature profile. A monotonically-decreasing pressure-temperature (P-T) profile would result in absorption rather than emission lines.

Here, we aim to detect or constrain which species may be responsible for this inversion. \cite{Yan22} already obtained non-detections for TiO, VO, and FeH in their CARMENES data, while \cite{Borsa22} did the same for AlO, TiO, and VO with HARPS-N; however, both of those spectrographs are fed by 4-m-class telescopes, motivating an independent search with 8-m-class telescopes and including additional potential inversion agents.

The PEPSI-LBT Exoplanet Transit Survey (PETS) is a large ($\sim$400 hour) survey of transiting exoplanets using the PEPSI high-resolution spectrograph on the Large Binocular Telescope (LBT). The first paper resulting from the survey set limits on the atmospheric presence of atomic species that could be vaporized from the crust of the super-Earth 55 Cnc~e \citep{PETS-I}. In this second paper in the series, we analyze our emission and transmission spectra of \thisplanet.

\section{Observations}

We obtained our data on \thisplanet\ with the Potsdam \'Echelle Polarimetric and Specrographic Instrument \citep[PEPSI;][]{Strassmeier15} on the $2\times8.4$ m Large Binocular Telescope (LBT) located on Mt. Graham, Arizona, USA. We observed using the 1.5'' fiber, giving a resolving power of $R=130,000$. We used Cross Dispersers (CD) III and V, covering 4800-5441 \AA\ and 6278-7419 \AA\ in the blue and red arms of the spectrograph, respectively. The 2019 transit observations were unbinned and used 600 second exposures, while the 2021 emission observations used $2\times1$ binning on the chip and 300 second exposures. 

\begin{figure}
\centering
\includegraphics[width=1.0\columnwidth]{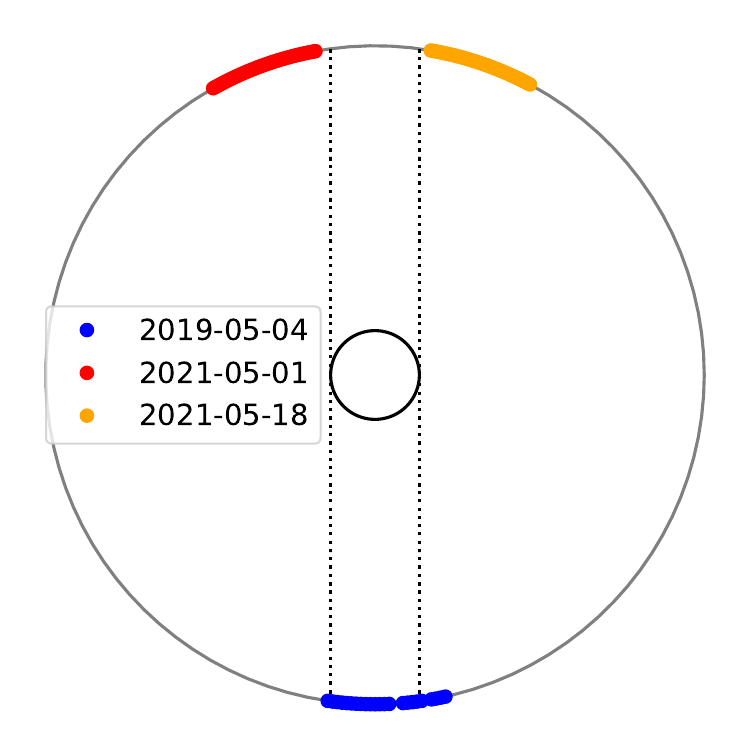}
\caption{Graphic illustrating the phases at which our observations of \thisplanet\ took place. Different nights are marked by different colors. The planetary orbit and the stellar surface are the large and small circles, respectively. The observer is viewing from the bottom, and the illustration is looking down on the north pole of the star, such that the planet moves in a counter-clockwise direction; transit occurs at the bottom, and secondary eclipse at the top. The dotted lines denote the edges of the secondary eclipse and the transit.}
\label{fig:phases}
\end{figure}

We observed one transit of \thisplanet\, and two $\sim4.5$-hour segments near secondary eclipse. We list the key parameters of our datasets in Table~\ref{tab:observations}, and show the phases covered by our observations graphically in Fig.~\ref{fig:phases}.

We reduced the data using the Spectroscopic Data Systems (SDS) pipeline, as described in \cite{Ilyin2000} and \cite{Strassmeier18}. The pipeline performs bias subtraction and flat field correction, order tracing and extraction, cosmic ray correction, wavelength calibration, and normalization and combination of all orders in each arm (red versus blue). The pipeline also estimates the variance in each pixel, which we pipe through our analysis procedure described below.

\begin{deluxetable*}{cccccccc}
\tablecaption{Log of Observations\label{tab:observations}}
\tablewidth{0pt}
\tablehead{
Date (UT) & Type & $N_{\mathrm{spec}}$ & Exp. Time (s) & Airmass Range & Phases Covered & SNR$_{\mathrm{blue}}$ & SNR$_{\mathrm{red}}$
}
\startdata
2019 May 4 & Transmission & 23 & 600 & 1.00-2.01 & -0.023-0.034 & 288 & 308 \\
2021 May 1 & Emission & 47 & 300 &  1.01-2.03 & 0.529-0.582 & 301 & 340 \\
2021 May 18 & Emission & 45 & 300 & 1.00-1.48 & 0.422-0.473 & 347 & 397 \\
\enddata
\tablecomments{$N_{\mathrm{spec}}$ is the number of spectra obtained on that night. Exp. time is the exposure time in seconds. SNR$_{\mathrm{blue}}$ and SNR$_{\mathrm{red}}$ are the nightly average of the $95^{\mathrm{th}}$ quantile per-pixel signal-to-noise ratios in the blue and red arms, respectively.} 
\end{deluxetable*}

\section{Analysis}

\subsection{Overall Procedure}
\label{sec:methods}

\begin{figure}
\centering
\includegraphics[width=1.0\columnwidth]{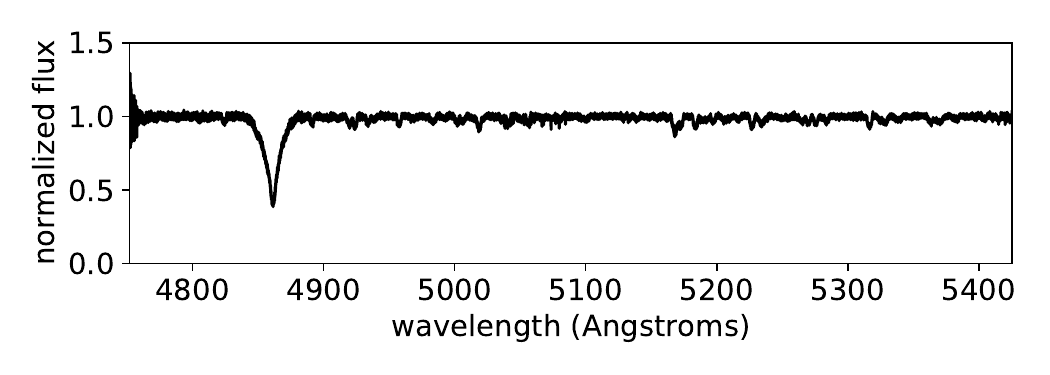}\\
\includegraphics[width=1.0\columnwidth]{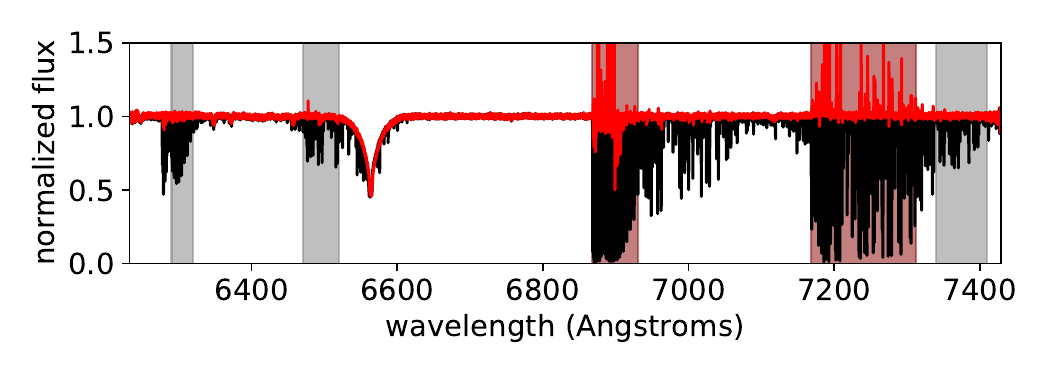} 
\caption{Example PEPSI data displaying our procedure to telluric-correct the spectra. The figure shows one PEPSI spectrum from 2021-05-01 in black, with the blue and red arms in the top and bottom panels, respectively. The \texttt{molecfit}-corrected spectrum is shown in red for the red arm. For that arm we also highlight the wavelength regions fit by \texttt{molecfit} in gray, and those which we mask out when performing our analyses due to poor telluric correction in red.} 
\label{fig:spectra}
\end{figure}

After receiving the reduced and normalized PEPSI spectra, more steps are necessary in order to detect the atmospheric signal. Our overall procedure is inspired by that of \cite{Nugroho17}, although with significant differences. We first run the \texttt{molecfit} package \citep{Smette15,Kausch15} on the red arm of each PEPSI spectrum to model out the telluric lines, as is described in more detail in \S\ref{sec:testing}; no such treatment is necessary for the largely telluric-free blue arm. We median-stack the telluric-corrected spectra and subtract this median spectrum from each of the time-series spectra in order to remove the stellar lines and the time-invariant component of the telluric lines. We then run the SYSREM algorithm \citep{Tamuz05} on these residuals, as described in \S\ref{sec:testing}. Next, we cross-correlate a model spectrum generated as described in \S\ref{sec:models} with each spectrum. The cross-correlation algorithm is adapted from that in the \texttt{BANZAI-NRES} package\footnote{\url{https://github.com/LCOGT/banzai-nres}}. These procedures, and the tests that we conducted to show that this is the optimal method, are detailed in \S\ref{sec:testing}. We show the spectra and the results of the \texttt{molecfit} 
in Fig.~\ref{fig:spectra}.

Finally, for a grid of possible radial velocity (RV) semi-amplitude $K_P$ values for the planet, we shift each of the time-series cross-correlation functions to remove the expected Doppler shift from that $K_P$ at the observation time; that is, we shift the spectra into the planetary rest frame. We assumed the ephemeris from \cite{Lund17}. We then stack the resulting shifted CCFs. For the transmission data we only combine the CCFs derived from the in-transit spectra. In this step we combine the CCFs derived from the blue and red arm data, and data taken on different nights. The exceptions are for FeH, where there are no lines present in the blue arm so we use only the red arm, and CaH, where we use only the blue arm due to the lack of red lines. We combine the CCFs using a weighted sum, where the weights are the 95$^{\mathrm{th}}$ percentile SNR from that arm multiplied by the total equivalent width in the model emission or transmission lines within that bandpass. 

We search for a CCF peak near the expected values of $K_P$, $v_\mathrm{sys}$ in this space; we calculated $K_P$ given the system parameters from \cite{Lund17}, and $v_\mathrm{sys}$ was measured by the PEPSI pipeline. These parameters are listed in Table~\ref{tab:modelpars}. We measure the SNR of the peak (if any) present at the location expected from these known parameters of the planet. 
We estimate the significance of any peaks using the standard deviation of the points in the shifted and combined CCFs with $|v|>100$ km s$^{-1}$ from the expected location of the planetary signal.

\subsection{Model Spectra}
\label{sec:models}

\begin{figure}
\centering
\includegraphics[width=1.0\columnwidth]{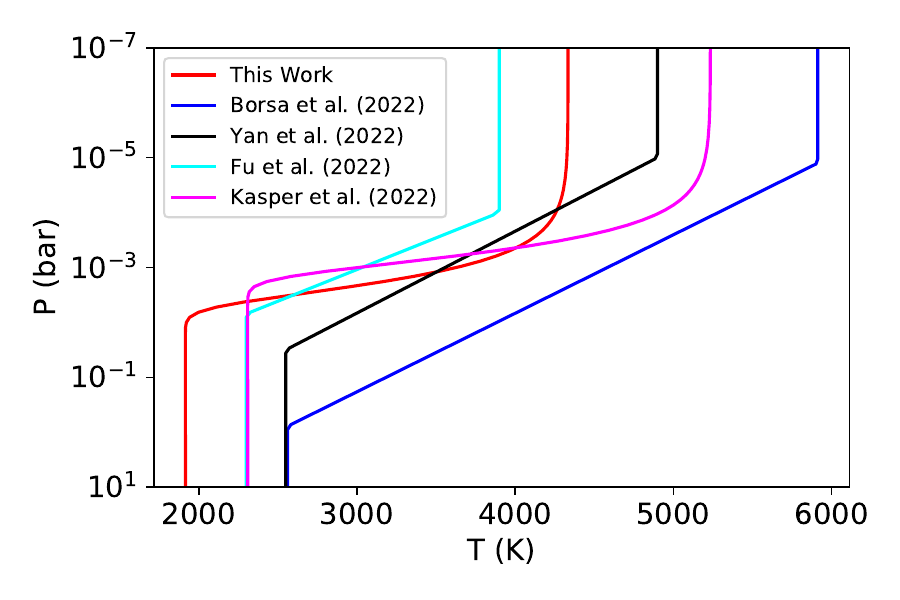}
\caption{Pressure-temperature profiles considered in this work. Our best P-T profile which we adopt is shown in red, along with the published P-T profiles retrieved by \cite{Yan22}, \cite{Fu22}, \cite{Borsa22}, and \cite{Kasper22}.
}
\label{fig:ptprofile}
\end{figure}

We generate model spectra using the \texttt{petitRADTRANS} package \citep{Molliere19}. We assumed the stellar and planetary parameters from \cite{Lund17}, which are listed in Table~\ref{tab:modelpars}. We assumed a P-T profile of the form given by Eqn.~29 of \cite{Guillot10} as implemented in \texttt{petitRADTRANS},
We tested this profile with a simple grid search in the parameters $\kappa_{\mathrm{IR}}$ and $\gamma$, along with the P-T profiles retrieved for \thisplanet\ by \cite{Yan22}, \cite{Fu22}, \cite{Borsa22}, and \cite{Kasper22}. We computed a model Fe~\textsc{i} spectrum for each profile assuming a VMR of $5.4\times10^{-5}$, cross-correlated it with the data as described below, and logged the SNR of the resulting detection. Our best model is a Guillot profile with $\gamma=30$, $\kappa_{\mathrm{IR}}=0.04$, which resulted in a $16.94\sigma$ detection. We adopt this P-T profile for the remainder of this work. None of the published retrieved profiles outperformed our model, the best being that from \cite{Borsa22} with $16.43\sigma$.
We show these profiles in Fig.~\ref{fig:ptprofile}, and list the results of the P-T profile comparison in Table~\ref{tab:ptprofiles}. We note that we used only an approximation to the P-T profile retrieved by \cite{Fu22}, as they used a more sophisticated profile tied to an atmospheric model. Instead, we approximated their profile using a simple model which is isothermal above and below the inversion, with a transition between the two regimes which is linear in log space, similar to those of \cite{Borsa22} and \cite{Yan22}. We also note that we only tested the best-fit P-T profile from each of the cited works and did not attempt to account for the quoted uncertainties on the profile parameters. For simplicity we use the same P-T profile for both the emission and transmission spectra; although in reality the P-T profile is likely to be different on the dayside versus the terminator, the dependence of the transmission spectrum on the details of the P-T profile is weak.

\begin{deluxetable}{cc}
\tablecaption{Pressure-Temperature Profile Comparison Results \label{tab:ptprofiles}}
\tablewidth{0pt}
\tablehead{
Source & Fe~\textsc{i} SNR ($\sigma$)
}
\startdata
Guillot $\gamma=30$, $\kappa_{\mathrm{IR}}=0.04$ (this work) & 16.94 \\
\cite{Borsa22} & 16.43 \\
\cite{Kasper22} & 16.06 \\
\cite{Yan22} & 13.97 \\
\cite{Fu22} & 12.98 \\
\enddata
\tablecomments{The Fe~\textsc{i} SNR is the SNR with which we measured the Fe~\textsc{i} cross-correlation signal assuming the specified P-T profile for a fixed Fe~\textsc{i} VMR of $5.4\times10^{-5}$.} 
\end{deluxetable}

\begin{deluxetable}{ccc}
\tablecaption{System and Atmospheric Model Parameters\label{tab:modelpars}}
\tablewidth{0pt}
\tablehead{
Parameter & Value & Source \\ 
}
\startdata
$R_P$ ($R_{\oplus}$) & 19.51  & \cite{Lund17} \\
$M_P$ ($M_{\oplus}$) & 1072  & \cite{Lund17} \\
$T_{\mathrm{eq}}$ (K) & 2262  & \cite{Lund17} \\
$P$ (d) & 3.4741085 & \cite{Lund17} \\
$T_0$ (BJD\_TDB) & 2457485.74965 & \cite{Lund17} \\
$R_{\star}$ ($R_{\odot}$) & 1.565  & \cite{Lund17} \\
$M_{\star}$ ($M_{\odot}$) & 1.76 & \cite{Lund17} \\
$T_{\mathrm{eff}}$ (K) & 8720  & \cite{Lund17} \\
$[$Fe/H$]$ & -0.29 & \cite{Lund17} \\
$K_P$ (km s$^{-1}$) & $169 \pm 6$ & \cite{Lund17} \\
$v_{\mathrm{sys}}$ (km s$^{-1}$) & $-26.0$ & \ldots \\
$v\sin i_P$ (km s$^{-1}$) & $2.5$ & \cite{Lund17} \\
$\kappa_{\mathrm{IR}}$ & 0.04 & \ldots \\
$\gamma$ & 30 & \ldots \\
$P_0$ (bar) & 1 & \ldots \\
$X_{\mathrm{H}_2}$ & 0.7496 & \ldots \\
$X_{\mathrm{He}}$ & 0.2504 & \ldots \\
VMR(H$^-$) & $1\times10^{-9}$ & \ldots \\
\enddata
\tablecomments{}
\end{deluxetable}

We generate spectra over the wavelength range 3850-7500 \AA, covering the full PEPSI CD III+V bandpass, with an even spacing of 0.01 \AA. We set up a model atmosphere with 130 grid points spanning pressures from 10$^2$ to 10$^{-10}$ bar.

We include continuum opacity due to $H^-$ bound-free absorption, H$_2$-H$_2$ and H$_2$-He collisions, and Rayleigh scattering due to H$_2$ and He. We assume a cloud-free atmosphere. We conservatively assumed an H$^-$ volume mixing ratio (VMR) of $1\times10^{-9}$, similar to that found by \cite{Arcangeli18} for WASP-18~b, a planet several hundred K hotter than \thisplanet. This is the major source of continuum opacity in the optical in our models, and is degenerate with the VMR ratio limits that we set using our transmission spectra (\S\ref{sec:injectionrecovery}; \S\ref{sec:results:agents}). Including this level of H$^-$ absorption worsens the transmission VMR limits by approximately two orders of magnitude over no H$^-$ absorption; this has no significant effect on the emission results. We also note that we assume a reference pressure level $P_0$ of 1 bar; changing $P_0$ to 1 mbar has a similar effect upon the VMR limits to including the noted H$^-$ absorption. We conservatively assume the high H$^-$ abundance in order to account for the unknown $P_0$ level and H$^-$ abundance. 
The single line absorber in question is the only line opacity source we include in our models. 

After computation of the model spectra, we convolve the models with a rotational broadening kernel and an instrumental broadening kernel. For the emission spectra we assume an analytic form for the rotational broadening kernel from Eqn.~18.14 of \cite{Gray}. For the transmission spectra we instead assume the form from Eqn.~15 of \cite{Gandhi22}, which results in a double-peaked profile from the limbs of the planet. We further assume that \thisplanet\ is tidally locked; given the planetary parameters from \cite{Lund17}, this implies an equatorial velocity of 2.5 km s$^{-1}$. For the instrumental broadening, we assume a Gaussian line spread function with a width corresponding to the PEPSI resolving power of $R=130,000$ \citep{Strassmeier18}.

The treatment of the emission versus transmission spectra is slightly different. For the emission spectra,
after calculating the planetary flux using \texttt{petitRADTRANS}, we generate a blackbody spectrum corresponding to the stellar $T_{\mathrm{eff}}$ and use this to estimate the planet-to-star flux ratio $F_P/F_{\star}$. It is this quantity that we use as as our model spectrum for the cross-correlation. 

For the transmission spectra, \texttt{petitRADTRANS} calculates the planetary radius as a function of wavelength $R_P(\lambda)$. As our spectra have been normalized by the data reduction pipeline and are not flux-calibrated, we do not have any information in our data on the planetary radius, only the differential change in radius as a function of wavelength. We therefore estimate the transmission spectrum as:
\begin{equation}
    F(\lambda) = 1-\frac{R_P(\lambda)^2-R_{P0}^2}{R_{\star}^2}
\end{equation}
where $R_{P0}$ is the broadband planetary radius, which we adopt for \cite{Lund17}.

We list the species considered and the sources of the opacity data used in Table~\ref{tab:opacities}.  We chose to search for the ``classical'' inversion agents TiO and VO \citep{Fortney08}, and the metal hydrides FeH and CaH. \texttt{petitRADTRANS} does not currently have high-resolution opacity tables for several other species of interest, such as AlO, CaO, NaH, or MgH \citep{GandhiMadhusudhan19}. In the interests of simplicity, for TiO we only use the single isotope $^{48}$TiO, as it makes up the majority of Ti atoms, and ignore the other stable isotopes. Likewise, for FeH the opacity table only considers the most common isotopologue, but unlike for TiO there are no \texttt{petitRADTRANS} opacity tables available for the other isotopologues.

\begin{deluxetable}{ccc}
\tablecaption{Opacity Data Sources\label{tab:opacities}}
\tablewidth{0pt}
\tablehead{
Species & Reference 1 & Reference 2 \\ 
}
\startdata
TiO & \cite{McKemmish19} & B. Plez (priv. comm.) \\
VO & \cite{McKemmish16} & B. Plez (priv. comm.) \\
FeH & \cite{Wende10} & \ldots \\
CaH & \cite{Li12} & \ldots \\
Fe~\textsc{i} & Kurucz & \ldots \\
\enddata
\tablecomments{Reference 1 is the opacity table used to generate the CCF template, and Reference 2 is that used to generate the spectrum injected for injection-recovery testing, if different. The opacity data for all species were obtained from the \texttt{petitRADTRANS} website\footnote{\url{https://petitradtrans.readthedocs.io/en/latest/content/available_opacities.html}}} The reference Kurucz refers to data from \url{http://kurucz.harvard.edu/}.
\end{deluxetable}

\begin{figure*}
\centering
\includegraphics[width=0.9\textwidth]{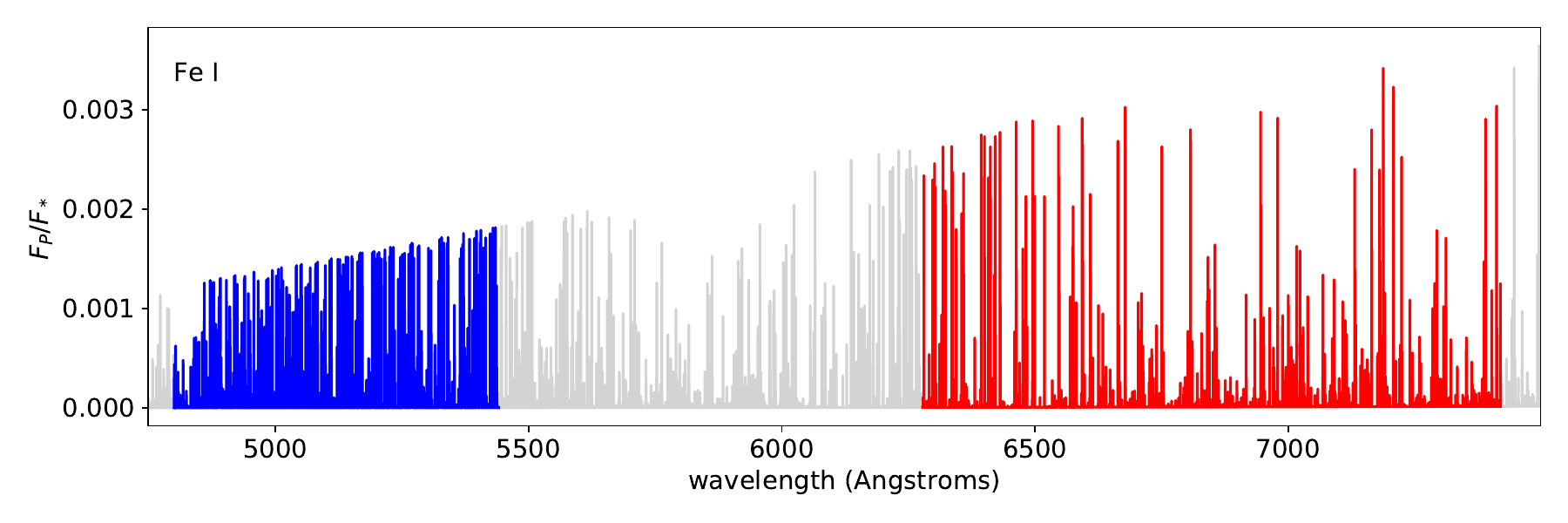} \\
\includegraphics[width=0.45\textwidth]{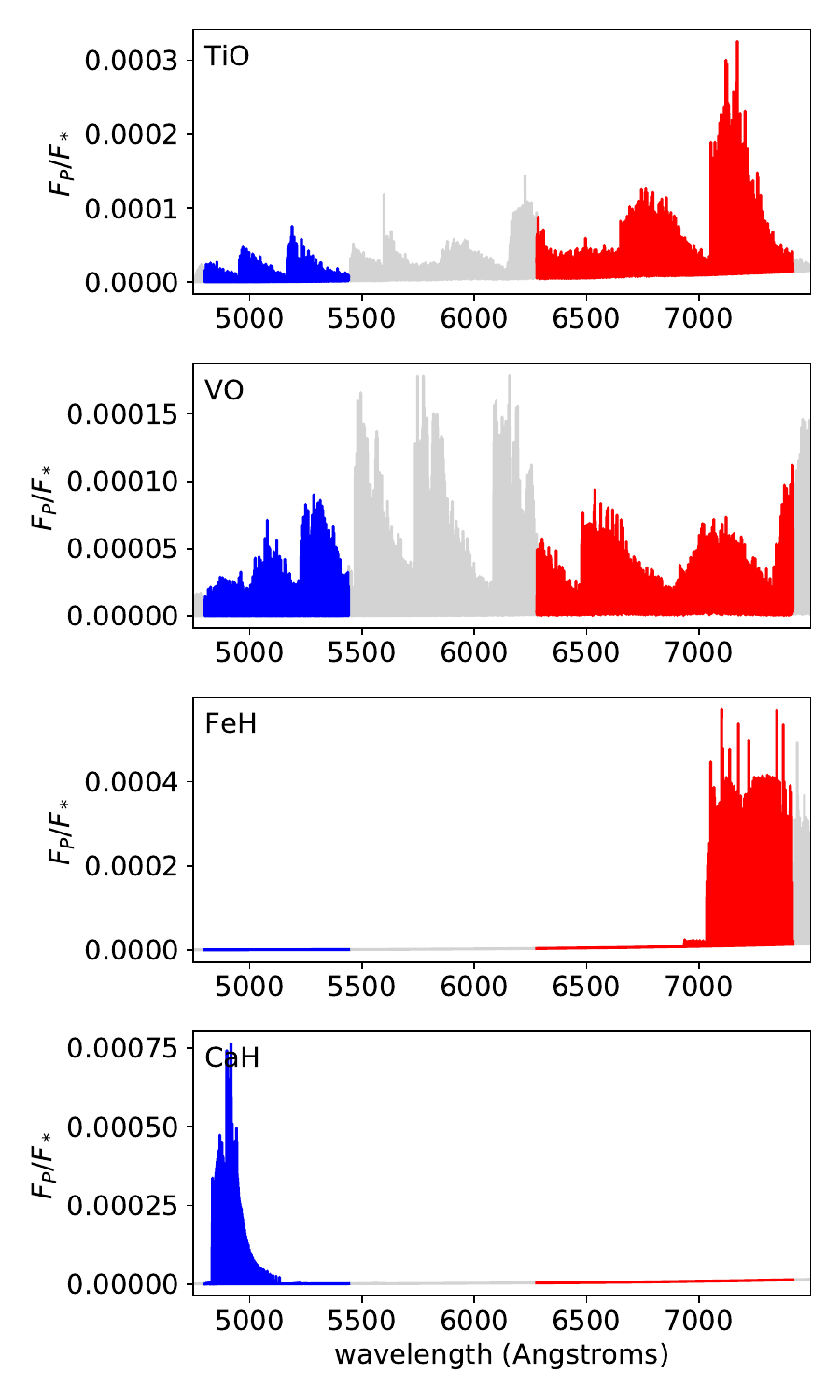}
\includegraphics[width=0.45\textwidth]{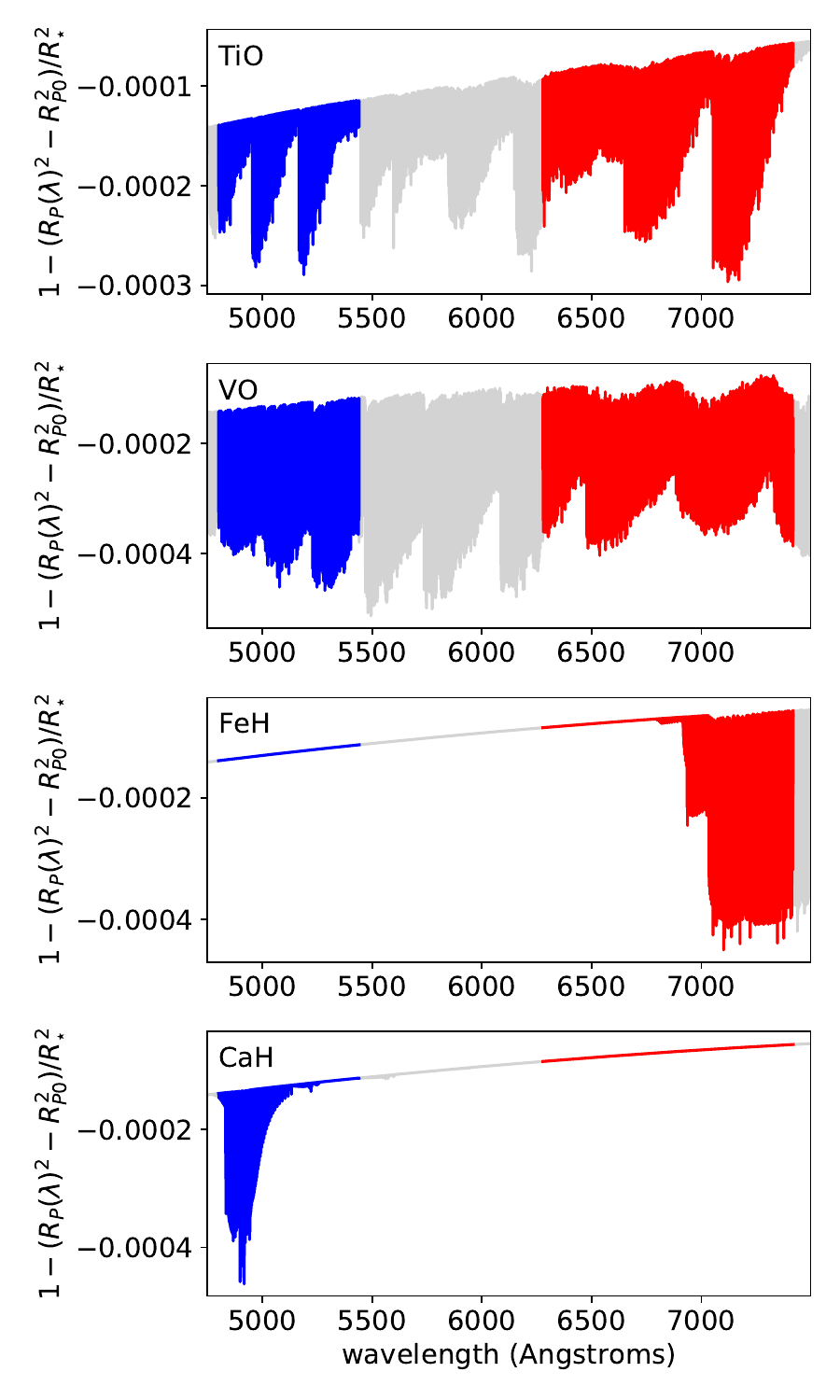}
\caption{Model spectra generated with \texttt{petitRADTRANS}. The wavelength ranges covered by PEPSI CD III and V are shown in blue and red, respectively. The top panel shows the emission spectrum of Fe~\textsc{i}. The left and right bottom columns show the emission and transmission spectra, respectively, of (top to bottom) TiO, VO, FeH, and CaH. The spectra shown for the molecular constituents are generated with the minimum VMR which would give a 4$\sigma$ detection in our injection-recovery testing; see the main text for more details.} 
\label{fig:modelspectra}
\end{figure*}

\subsection{Comparison of Telluric and Systematics Correction Methods}
\label{sec:testing}

In order to detect the minute planetary signal, it is necessary to correct for systematic sources of error, most notably the presence of telluric absorption lines in the spectra, but also systematics due to detector imperfections, etc. 
Most recent studies of exoplanetary atmospheres at high resolution use one of two methods to correct systematics: SYSREM \citep[e.g.,][]{Nugroho20,Merritt21,Cont21} or \texttt{molecfit} \citep[e.g.,][]{Stangret21,Cabot21,Kesseli22}.  The SYSREM algorithm \citep{Tamuz05} is a PCA-based algorithm to remove common linear or time-varying systematics. The \texttt{molecfit} package \citep{Smette15,Kausch15} fits and removes the telluric lines in the spectrum. After implementing both methods, we performed tests to evaluate which method works best for our PEPSI data.

In order to implement SYSREM, we modified the source code from the \texttt{PySysRem} package\footnote{https://github.com/stephtdouglas/PySysRem} to handle the PEPSI data. We added code to formally propagate uncertainties through the SYSREM algorithm. For all runs, we used the airmass of the exposure as the initial guess for the starting values of the first systematic removed. For the red arm, we split the spectra into regions with telluric contamination versus no telluric contamination, with a threshold of $>1\%$ absorption over multiple nearby lines, varying the number of systematics removed for the former regions and removing one systematic from the latter. Our tests indicated that the recovered SNR did not change significantly if we removed two versus one systematics from the telluric-free regions.

Although \texttt{molecfit} was designed to work on data from ESO facilities, it has also been used for data from other spectrographs \citep[e.g.,][]{Cabot21}. We only used \texttt{molecfit} to correct the PEPSI red arm data, as there are no significant telluric lines in the blue arm. We fit three narrow regions spanning the red arm (6290-6320 \AA, 6470-6520 \AA, and 7340-7410 \AA) containing moderately-strong telluric lines, and included H$_2$O and O$_2$ in our models. While the telluric model removes weak lines adequately, it does not fit the strong lines to any good precision as they are saturated, and we consequently mask out regions of the spectra containing poorly-corrected lines in our further analysis. The telluric correction is sufficiently good for weak lines that any remaining residuals can be removed well by SYSREM (see below).

For all our tests, we inject a model emission spectrum of TiO, VO, FeH, or CaH into the PEPSI data. We constructed the model spectra as described in \S\ref{sec:models}, using VMRs chosen to give a moderate but not overwhelming detection significance. We passed these spectra through SYSREM and then attempted to recover the signal using the cross-correlation methodology described in \S\ref{sec:methods}. We evaluated different numbers of SYSREM systematics removed in order to find the optimal number. For the red arm, we evaluated using SYSREM alone, \texttt{molecfit} alone, or \texttt{molecfit} followed by SYSREM. 

\begin{figure*}
\centering
\includegraphics[width=0.45\textwidth]{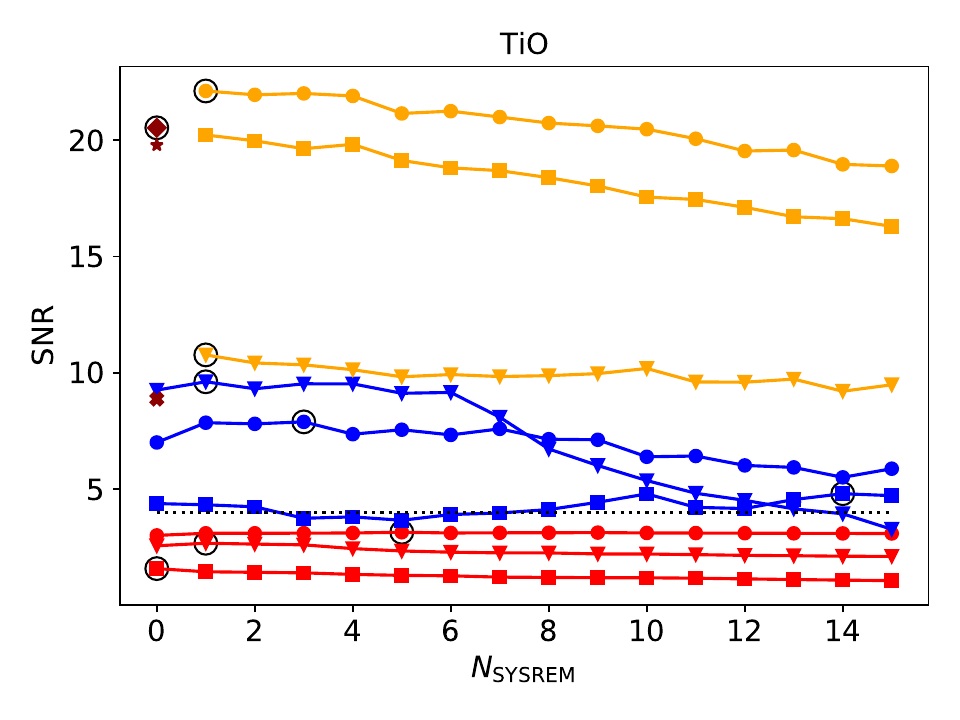}
\includegraphics[width=0.45\textwidth]{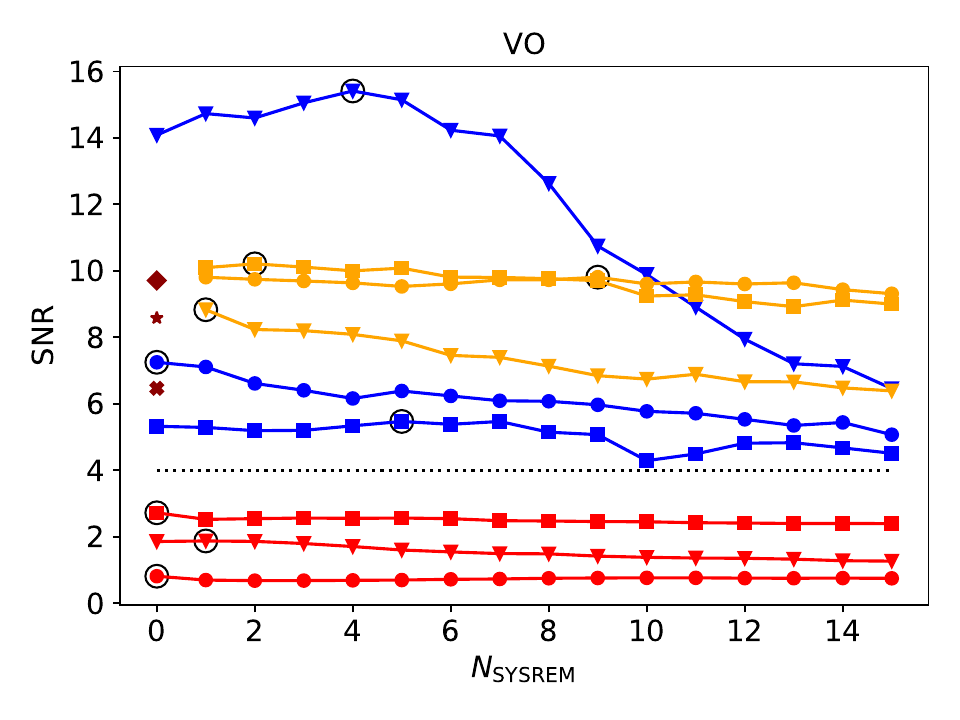}\\
\includegraphics[width=0.45\textwidth]{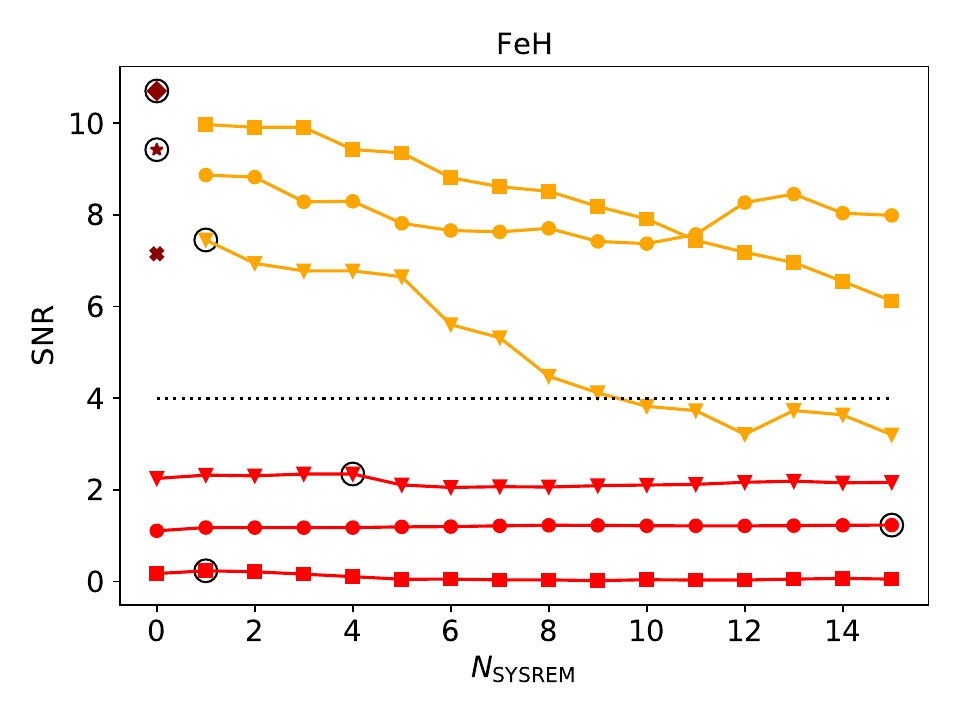}
\includegraphics[width=0.45\textwidth]{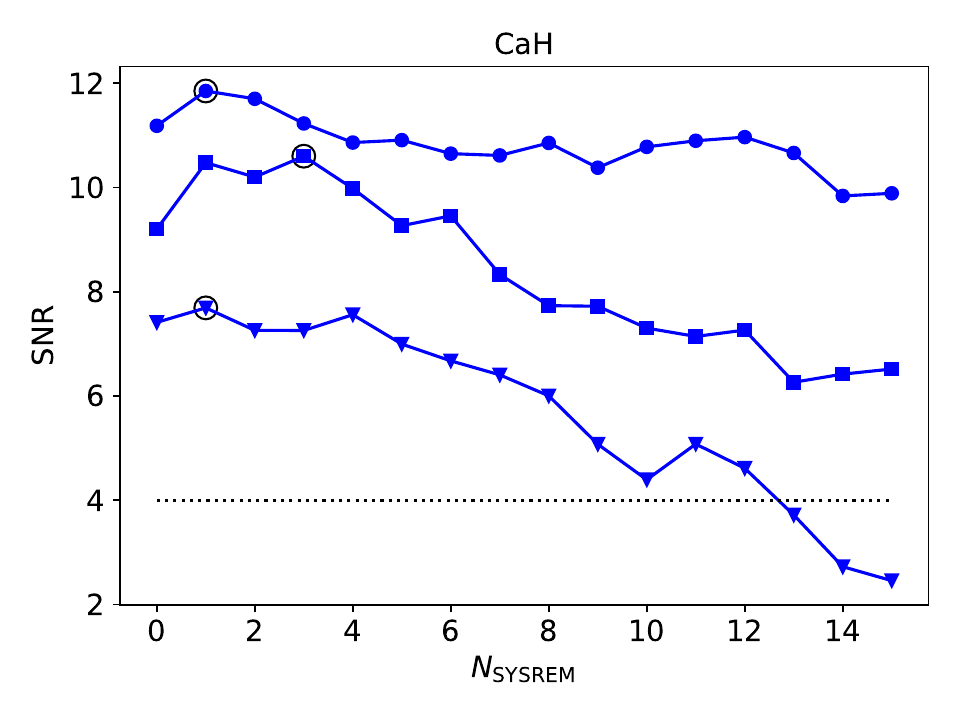}\\
\caption{Results of the tests of the systematics-correction methods. Lines of the same color show the same method and PEPSI arm: blue arm SYSREM (blue), red arm SYSREM (red), red arm \texttt{molecfit} (dark red), and red arm \texttt{molecfit} + SYSREM (orange). Circles and stars show the 2021-05-01 dataset, squares and diamonds the 2021-05-18 dataset, and triangles and x's the 2019-05-01 dataset; the latter dataset is the transmission data. The x-axis shows the number of systematics that have been removed with SYSREM.The plots show the results for TiO (upper left), VO (upper right), FeH (lower left), and CaH (lower right). Injections resulting in peaks below 4$\sigma$ (the horizontal dashed line) are not considered to be recovered but are shown for completeness. The arm/method combination resulting in the highest SNR for each dataset is highlighted with a black circle. In all cases for the red arm using \texttt{molecfit}-only outperforms using SYSREM only, and in many cases using both offers further improvements.
}
\label{fig:snrcomparison}
\end{figure*}

We show the results of these tests in Fig.~\ref{fig:snrcomparison}. 
In all cases for the red arm, using \texttt{molecfit} alone outperforms using SYSREM alone. In many cases, the combination of \texttt{molecfit} and SYSREM outperforms using either alone. For most of the datasets, the optimal number of systematics removed is small (1-4). In only two of the eighteen datasets considered (arms $\times$ nights) is the optimal number greater than 5.
We summarize the optimal number of systematics removed that we use for the remainder of the paper in Table~\ref{tab:sysrem}.

For the blue arm, the optimal number of SYSREM iterations is generally $\leq5$, except for one TIO dataset which attained a maximum SNR after 14 iterations.

We did not inject the FeH signal in the blue arm because it does not have any significant lines in that region of the spectrum, and similarly we did not inject the CaH signal into the red arm.

Overall, the optimal strategy that we have used as is follows. For TiO and VO, we use both \texttt{molecfit} and SYSREM. For FeH, we use only \texttt{molecfit}, while for CaH, we use only SYSREM, since its lines are only present in the blue arm. For each species we adopt a number of SYSREM iterations approximating the average of the best number for the three datasets, which we list in Table~\ref{tab:sysrem}.

\begin{deluxetable}{cccccc}
\tablecaption{Optimal Number of SYSREM Iterations \label{tab:sysrem}}
\tablewidth{0pt}
\tablehead{
Date & Arm & TiO & VO & FeH & CaH \\
}
\startdata
2019-05-04 & red & 1 & 1 & 1 & \ldots \\
2021-05-01 & red & 1 & 9 & 0 & \ldots \\
2021-05-18 & red & 0 & 2 & 0 & \ldots \\
Adopted & red & 1 & 2 & 0 & \ldots \\
2019-05-04 & blue & 1 & 4 & \ldots & 1 \\
2021-05-01 & blue & 3 & 0 & \ldots & 1 \\
2021-05-18 & blue & 14 & 5 & \ldots & 3 \\
Adopted & blue & 2 & 3 & \ldots & 2 \\
\enddata
\tablecomments{}.
\end{deluxetable}

\subsection{Injection-Recovery Testing}
\label{sec:injectionrecovery}

In cases where we obtain a non-detection, it is useful to attempt to quantify these non-detections. Do the data constrain the non-detection to an interesting level, or are the data simply not high enough quality to place useful limits? We attempt to address this issue through injecting a model spectrum into our data before treating the data with SYSREM, and then recovering it with our cross-correlation process described in \S\ref{sec:methods}. For each species, we inject models with VMRs spaced 0.25 dex apart, i.e., VMR=[1.0, 1.8, 3.2, 5.6]$\times10^x$, and find the smallest VMR for which we can recover the signal at $>4\sigma$  \citep[cf.][]{Kesseli20}. We note that these results are contingent upon the choice of P-T profile, as the strength of the inversion affects the strength of the emission lines, and thus the inferred VMR limit. However, \cite{Yan22}, \cite{Fu22}, \cite{Borsa22}, and \cite{Kasper22} all independently retrieved broadly similar P-T profiles using independent methodology and data, giving us some confidence in the broad accuracy of our P-T profile. Additionally, the presence of additional or stronger continuum opacity sources than we have assumed could affect the limits. These results should therefore only be taken as approximate under the assumption of the P-T profile and other model atmospheric parameters.

One other potential issue is with the inaccuracy of line lists, which can cause systematic reductions in signals strong enough to result in a non-detection when a detection would otherwise be expected \citep[e.g.,][]{deRegt22}. In order to attempt to account for this issue, for species with multiple opacity tables derived from different line lists available (i.e., TiO and VO), we generate separate model spectra using the two line lists, inject one into the data, and attempt to recover it using the other line list. We have found that this worsens the detectable VMR by approximately 1 order of magnitude for TiO. For VO, on the other hand, this worsens the limits by two orders of magnitude for the emission spectra and the transmission spectra are not detectable at any VMR when using a different template due to their lower SNR. For completeness we quote the VO limits using both the same and differing line lists.

\section{Results}

\subsection{Recovery of Previous Detections}

\begin{figure}
\centering
\includegraphics[width=1.0\columnwidth]{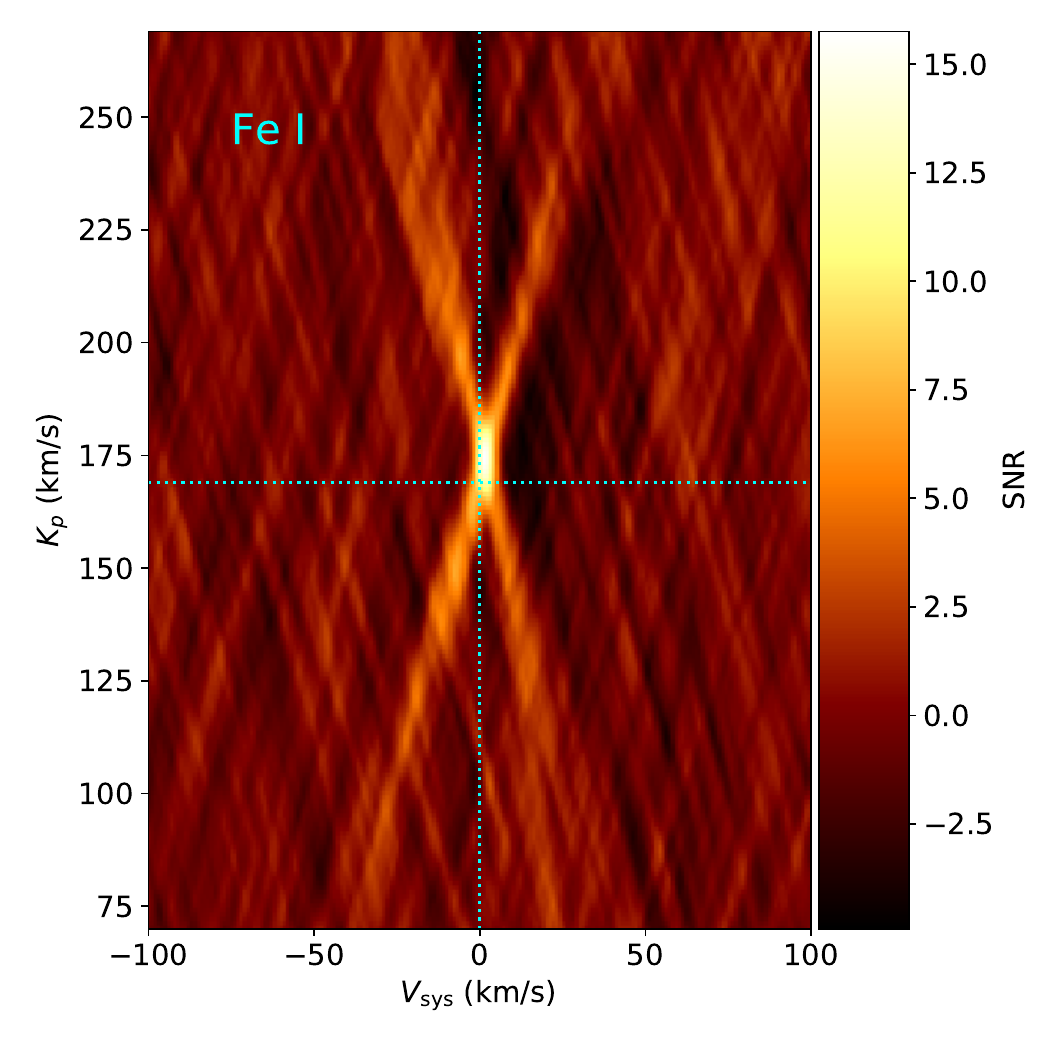}
\caption{Shifted and combined CCFs for emission from Fe~\textsc{i}. We detect the Fe~\textsc{i} emission at a significance of $16.9\sigma$, reproducing the results of \cite{Yan22}, \cite{Borsa22}, and \cite{Kasper22}.}
\label{fig:ccfs-fesicr}
\end{figure}

In order to perform a final validation of our methodology and data, we attempt to reproduce the detections of emission from Fe~\textsc{i} from \cite{Yan22}, \cite{Borsa22}, and \cite{Kasper22}. We estimated VMRs for this model using
a \texttt{FastChem} \citep{Stock18} equilibrium chemical model as implemented in \texttt{PyFastChem}\footnote{\url{https://github.com/exoclime/FastChem}}. We assumed a solar abundance mixture, and the same P-T profile and pressure grid that we used to construct our model spectra.
We assumed the maximum VMRs present in the \texttt{FastChem} models, $5.4\times10^{-5}$ for Fe~\textsc{i}. 
We show the model spectra in the left panels of Fig.~\ref{fig:modelspectra}, and the CCFs shifted and stacked according to the expected planetary orbit in Fig.~\ref{fig:ccfs-fesicr}.

We were able to recover a signal from Fe~\textsc{i} near the expected $K_P$, $v$ values at a significance of $15.1\sigma$. \cite{Yan22} obtained a $7.7\sigma$ detection, and \cite{Borsa22} at $7.1\sigma$. It is unsurprising that we obtain a stronger detection, due to the much larger aperture of LBT ($2\times8$ m) versus the smaller telescopes used by those works (3.5 m at Calar Alto Observatory, and 3.58 m at the Telescopio Nazionale Galileo), although our bandpass is smaller than that of these instruments, partially offsetting this advantage.

\subsection{Limits on Inversion Agents}
\label{sec:results:agents}

We searched for TiO, VO, FeH, and CaH emission and transmission in our spectra of \thisplanet, and did not make any detections with a signficance of $>4\sigma$. Although the TiO transmission data do contain a bump near the correct ($v$, $K_P$) peaking at $2.94\sigma$ (Fig.~\ref{fig:ccfs-tio}), due to the presence of other similarly-strong peaks elsewhere in the ($v$, $K_P$) space we do not consider this to be a robust detection. We show the data and minimum recovered peaks in the Appendix.

\begin{figure}
\centering
\includegraphics[width=1.0\columnwidth]{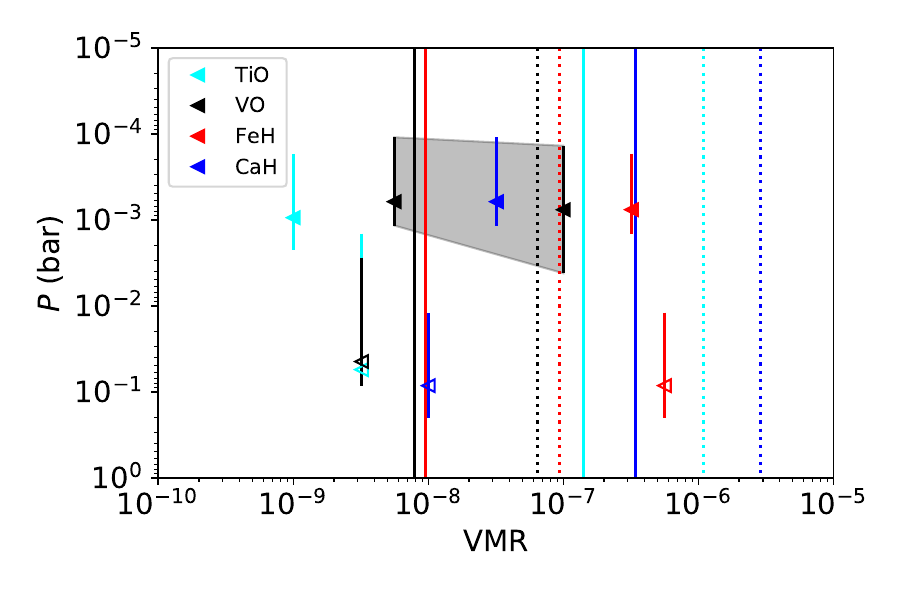}
\caption{Graphical summary of our limits on the abundances of inversion agents. For each species, the left-pointing arrow shows our upper limit set upon the VMR, while the error bar shows the range in altitudes which contribute 68\% of the emission or absorption from that species within the PEPSI bandpass at the VMR limit. The filled point for each species is from the emission data, and the empty points from the transmission data. The vertical lines show the VMRs expected from a chemical equilibrium calculation at 1 bar, assuming that this is the quenching pressure and that the VMR is constant above this point. Solid lines assume solar metallicity, and dotted lines $10\times$ solar. The colors of the lines correspond to those of the data points for each species. If the data point is to the left of the colored bar, the observation is in conflict with the expectation.}
\label{fig:results}
\end{figure}

Instead, we set limits upon the VMRs of these species as described in \S\ref{sec:injectionrecovery}. These limits range from $3.2\times10^{-7}$ for FeH to $1\times10^{-9}$ for TiO for the emission data, and from $5.6\times10^{-7}$ for FeH to $3.2\times10^{-9}$ for TiO for the transmission data. We list the full results in Table~\ref{tab:limits}, and show the results graphically in Fig.~\ref{fig:results}. 

We derive the pressure levels for the VMR limits shown in Fig.~\ref{fig:results} from the contribution functions calculated by \texttt{petitRADTRANS}; we calculate the pressure of peak contribution and show as the error bars the range of pressures over which $68\%$ of the contribution takes place. Counterintuitively, in our models the transmission spectra probe \textit{deeper} into the atmosphere than the emission spectra. This is because, in our models, the lack of other optical opacity sources besides the absorber in question allows the transmission rays to penetrate deep into the atmosphere, below the temperature inversion. The emission lines, on the other hand, by definition are formed only in the inversion. 
Although the nominal transmission and emission models are consistent with each other, they are unlikely to be completely realistic. We also calculated spectra with an artificially increased VMR of H$^-$ to $1\times10^{-9}$, similar to the maximum VMR found by \cite{Arcangeli18} for WASP-18~b, a planet $\sim600$ K hotter than \thisplanet. Including this artificially increased H$^-$ absorption does not significantly affect the emission results, but for the transmission results the limits worsen by 1-2 orders of magnitude, and the contribution moves by $\sim1$ dex lower in pressure. 

\begin{deluxetable*}{cccccc}
\tablecaption{Summary of Results\label{tab:limits}}
\tablewidth{0pt}
\tablehead{
Species & Emission VMR limit & Recovered SNR & Transmission VMR limit & Recovered SNR & Expected VMR 
}
\startdata
TiO & $1\times10^{-9}$  & 4.29 & $3.2\times10^{-9}$ & 4.30 & $1.4\times10^{-7}$ \\
VO (cross) & $1\times10^{-7}$ & 4.63 & \ldots & \ldots & $8.0\times10^{-9}$ \\
VO (self) & $5.6\times10^{-9}$ &  4.63 &  $1.8\times10^{-9}$ & 4.60 & $8.0\times10^{-9}$ \\
FeH & $3.2\times10^{-7}$ & 5.35 &  $5.6\times10^{-7}$ & 4.84 & $9.3\times10^{-9}$ \\
CaH & $3.2\times10^{-8}$ & 6.61 & $1\times10^{-8}$ &  4.02 & $3.4\times10^{-7}$ \\
\enddata
\tablecomments{The ``recovered SNR'' is the SNR of the CCF peak recovered for the injected model spectrum at the quoted VMR. For VO, we quote the results using both injecting and recovering a spectrum from the same line list (``self'') and differing line lists (``cross'').}.
\end{deluxetable*}

In order to interpret the results, we used a \texttt{FastChem} \citep{Stock18} equilibrium chemical model as implemented in \texttt{PyFastChem}\footnote{\url{https://github.com/exoclime/FastChem}}. We constructed two models, one assuming a solar abundance mixture and the other $10\times$ solar metallicity, with the same P-T profile and pressure grid that we used to construct our model spectra. We assumed that quenching occurs at a pressure of 1 bar, such that the VMRs of our species of interest are constant at $P<1$ bar \citep[cf.][]{ZahnleMarley14,MarleyRobinson15,Fortney20}. \texttt{FastChem} does not include photoionization or photodissociation, which may be important processes in the upper atmosphere of \thisplanet. Photodissociation should serve to depress the concentrations of molecules including inversion agents, while photoionization should increase the concentrations of ionized species and H$^-$. These effects are counteracted by the replenishing process of vertical mixing when we assume a quench pressure at 1 bar. Although these assumptions regarding quenching and the model's disregard of irradiation effects are likely to be only approximately correct, this nonetheless serves as a useful tool to help interpret our limits. 

We show these predicted VMRs from \texttt{FastChem} as vertical lines in Fig.~\ref{fig:results}. Our VMR limits for TiO and CaH lie several orders of magnitude below the expectations for those species at both metallicities. Our limits for FeH lie above the expected concentrations for all metallicities and are thus not constraining. Most of the strong FeH lines lie redward of the PEPSI CD V bandpass; new observations with PEPSI CD VI \citep[which covers 7419-9067 \AA;][]{Strassmeier15} would be required to better constrain FeH with PEPSI. For VO, our limits are constraining only if the ExoMol line list \cite{McKemmish16} is accurate; otherwise, we cannot claim a constraining limit (cf. \cite{deRegt22}). We again note that our VMR limits are only approximate, and in detail subject to our assumptions regarding the P-T profile, continuum opacity from H$^-$, and reference pressure level. Nonetheless, the large mismatch between the expected VMRs and our limits for TiO and CaH suggest that these species are depleted in the atmosphere of \thisplanet\ and cannot be responsible for the temperature inversion, while VO may also be depleted if the ExoMol line list is sufficiently accurate.

This could be due to cold trapping of these species by rain-out of condensed species on the cold planetary nightside \citep[e.g.,][]{Spiegel09,Parmentier13}, which was previously suggested to occur for \thisplanet\ by \cite{Nugroho20}. If these species can condense out into particles that are large enough to settle deep into the atmosphere despite any convection on the nightside, then these constituents will be depleted from the upper layers of the atmosphere that we can probe with transmission or emission spectroscopy. 
Neither atomic Ti nor V have been detected in the transmission spectrum of \thisplanet\ \citep{Nugroho20}. Since neither the expected atomic or molecular Ti- and V-bearing species are seen in the atmosphere of \thisplanet, it is plausible that rain-out has removed these molecules from the upper atmosphere.
Rain-out, however, cannot explain the lack of detections of CaH, as \cite{Nugroho20} and \cite{CasasayasBarris19} detected Ca in the transmission spectrum, indicating that this species do not rain out. Alternately, molecular species could be photodissociated by the intense ultraviolet radiation from the early-type host star \thisstar. \cite{Tsai21}, however, found that for the even hotter UHJ WASP-33~b that photochemistry should dominate only at high altitudes ($P<10^{-4}$ bar).

\section{Conclusions}

We presented observations of the atmosphere of \thisplanet\ with LBT/PEPSI in both emission and transmission. We recovered the previously-detected Fe~\textsc{i} emission from the planetary atmosphere at a significance of nearly $17\sigma$.

We set strict limits upon the presence of the possible inversion agents TiO, VO, FeH, and CaH in the atmosphere of \thisplanet\ (Fig.~\ref{fig:results}, Table~\ref{tab:limits}), although these limits are subject to our assumptions regarding the P-T profile and continuum opacity, and the accuracy of the VO line list. Nonetheless, together with expectations from simple chemical models, this suggests that TiO, CaH, and possibly VO are depleted in the atmosphere of \thisplanet\ and cannot be responsible for the temperature inversion. Either a different molecular agent is responsible for the inversion in \thisplanet, or, as suggested by \cite{Nugroho20}, atomic species such as Fe~\textsc{i} could be responsible. Alternately, if there is indeed significant H$^-$ opacity in the atmosphere of \thisplanet\, it could be the opacity responsible for the inversion \citep{Lothringer18}.

In this paper we have presented our results on \thisplanet. As part of the PETS survey we have already gathered similar observations of other UHJs, with the aim of conducting a population study of inversion agents, which we will present in a future paper along with improved analysis techniques.  These data will also be more broadly useful for characterizing these planets. Given the strength of our Fe~\textsc{i} detection, we may be able to use the time-resolved emission to constrain the night-side brightness and hotspot offset of \thisplanet\ much like \cite{Herman22} were able to do for WASP-33~b, but doing so is beyond the scope of the present paper and will be presented in a future work along with further analysis of the other atomic species.

\vspace{12pt}

Thanks to Wilson Cauley for technical assistance, and Francesco Borsa and Lorenzo Pino for useful discussions. We thank the anonymous referee for insightful comments which improved the quality of the paper.

B.S.G. was partially supported by a Thomas Jefferson Chair for Space Exploration endowment at the Ohio State University.
This research was supported by the Excellence Cluster ORIGINS which is funded by the Deutsche Forschungsgemeinschaft (DFG, German Research Foundation) under Germany's Excellence Strategy - EXC-2094 - 390783311.

The LBT is an international collaboration among institutions in the United States, Italy and Germany. LBT Corporation partners are: The Ohio State University, representing OSU, University of Notre Dame, University of Minnesota and University of Virginia; The University of Arizona on behalf of the Arizona Board of Regents; Istituto Nazionale di Astrofisica, Italy; and LBT Beteiligungsgesellschaft, Germany, representing the Max-Planck Society, The Leibniz Institute for Astrophysics Potsdam, and Heidelberg University.

\facility{LBT (PEPSI)}

\software{\texttt{astropy} \citep{exoplanet:astropy13, exoplanet:astropy18},
\texttt{jupyter} \citep{jupyter},
\texttt{matplotlib} \citep{matplotlib},
\texttt{molecfit} \citep{Smette15,Kausch15},
\texttt{numpy} \citep{numpy},
\texttt{petitRADTRANS} \citep{Molliere19},
\texttt{PyFastChem} \citep{Stock18},
    \texttt{uncertainties} \citep{uncertainties}
}

\bibliography{sample631}{}
\bibliographystyle{aasjournal}



\appendix

\section{Cross-Correlation Functions for Inversion Agents}

\begin{figure*}
\centering
\includegraphics[width=0.45\textwidth]{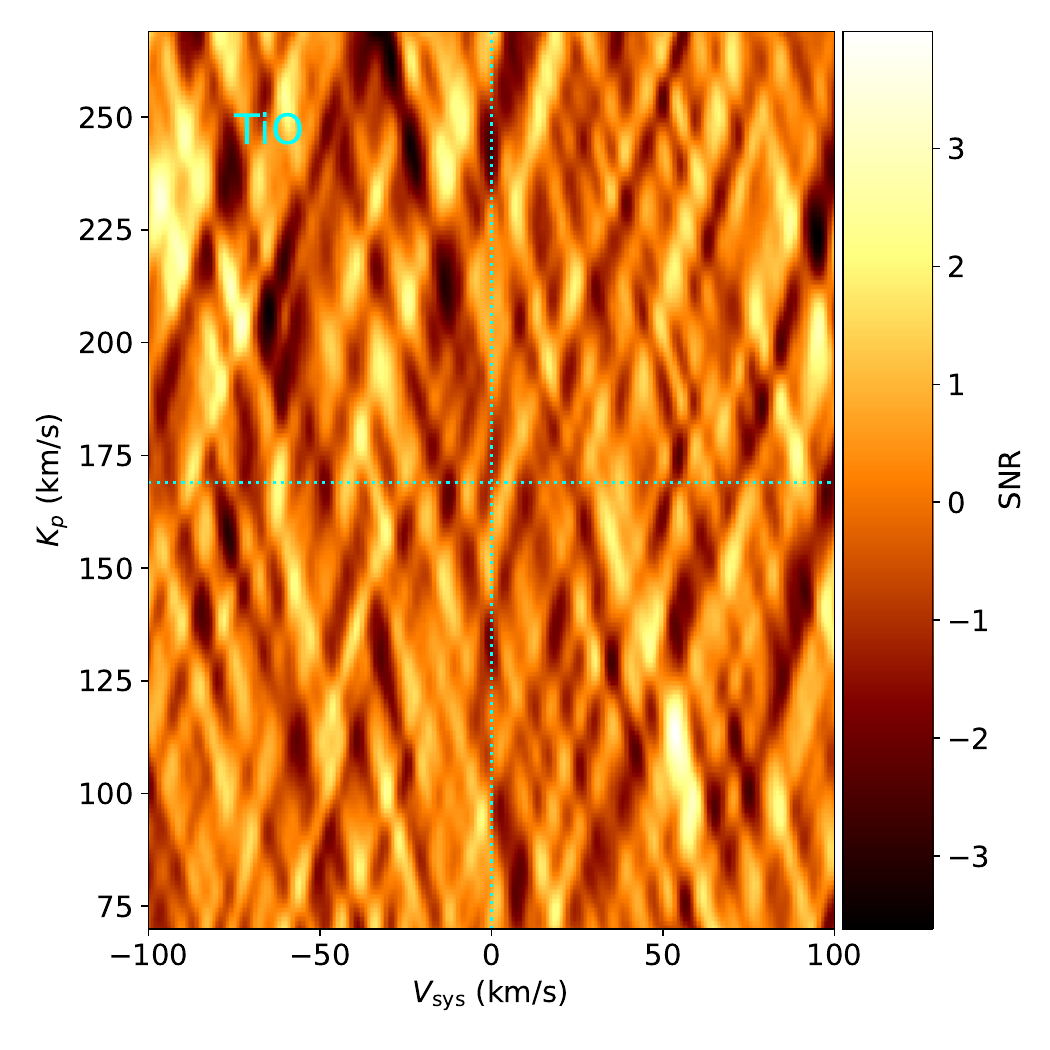}
\includegraphics[width=0.45\textwidth]{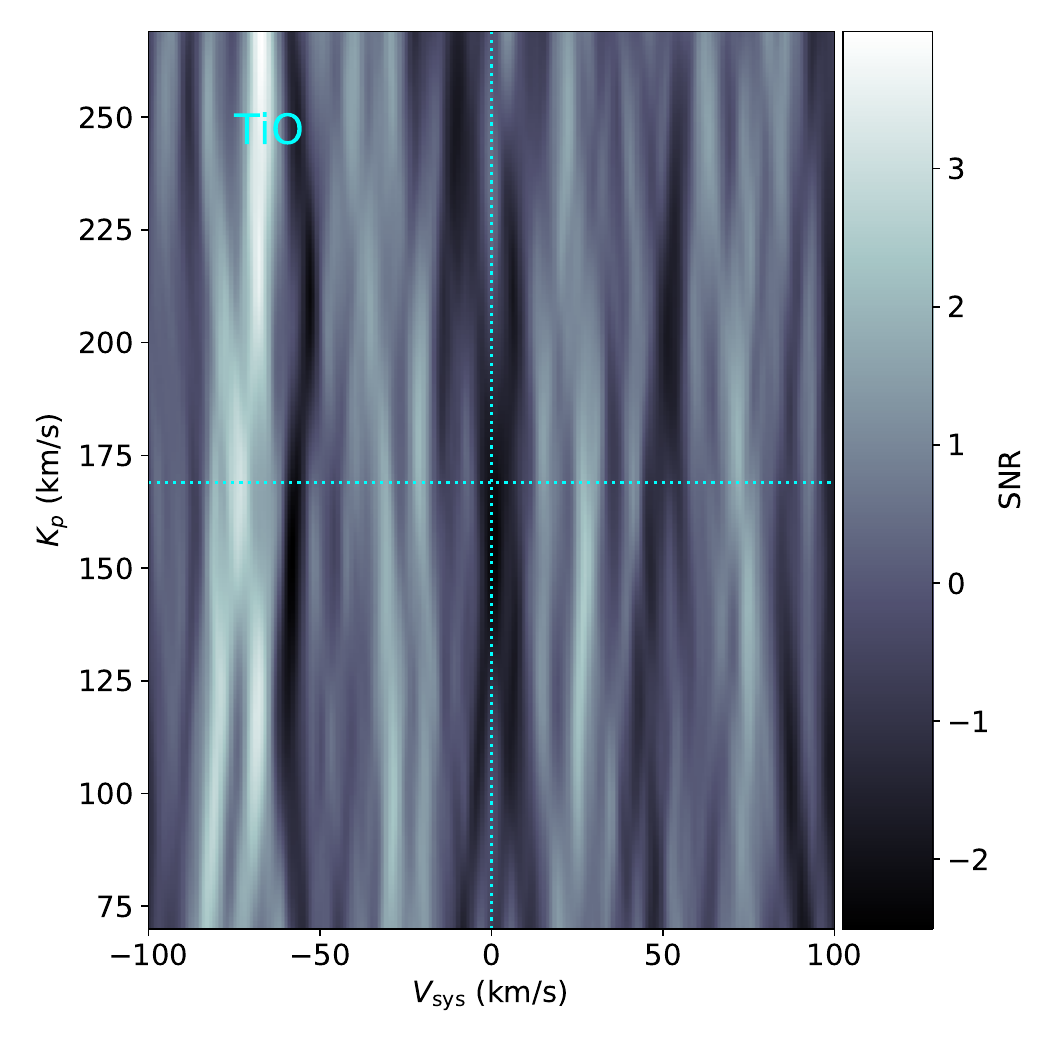}\\
\includegraphics[width=0.45\textwidth]{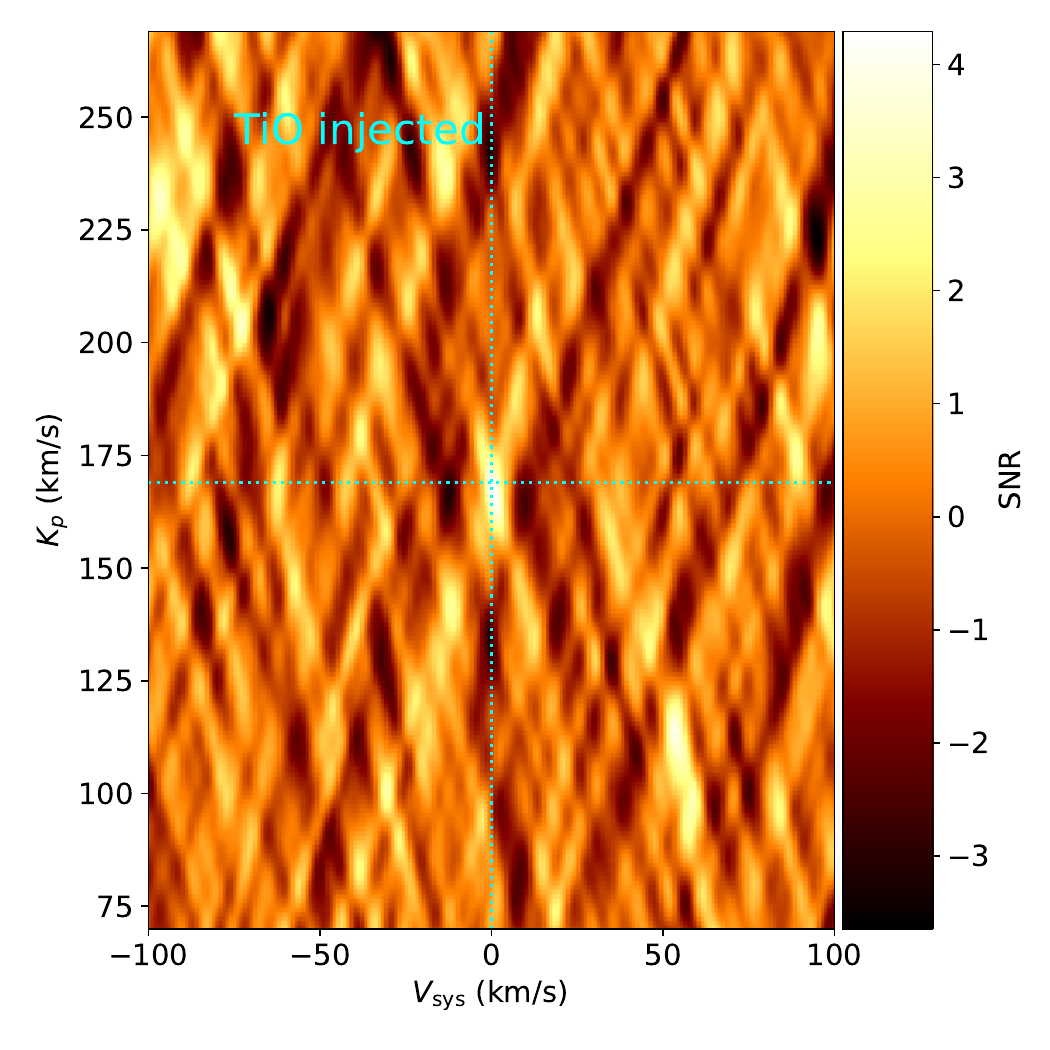}
\includegraphics[width=0.45\textwidth]{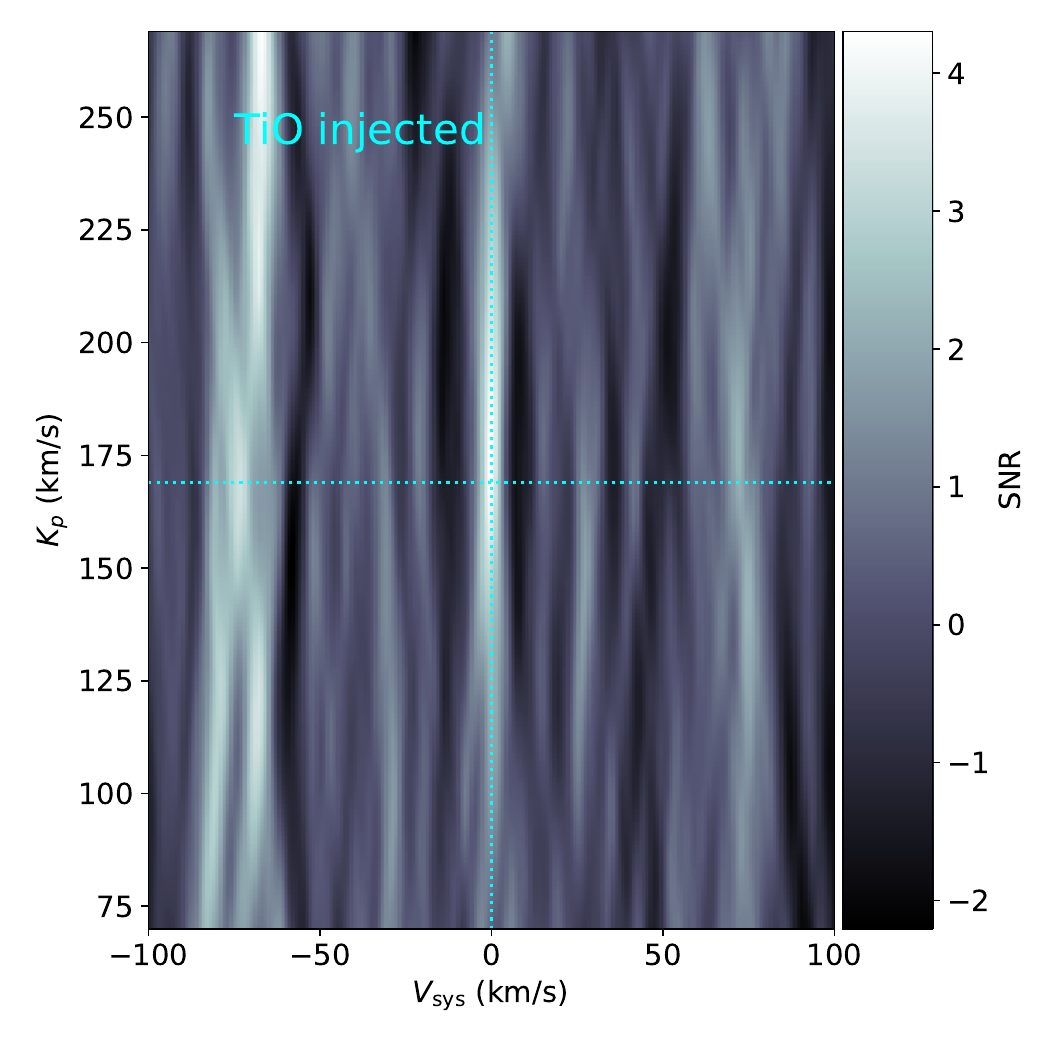}
\caption{Shifted and combined CCFs for TiO, in emission (left panels) and transmission (right panels). The top panels show the data, and the bottom panels the data with a signal from the minimum VMR that can be recovered at $>4\sigma$ injected.}
\label{fig:ccfs-tio}
\end{figure*}

\begin{figure*}
\centering
\includegraphics[width=0.45\textwidth]{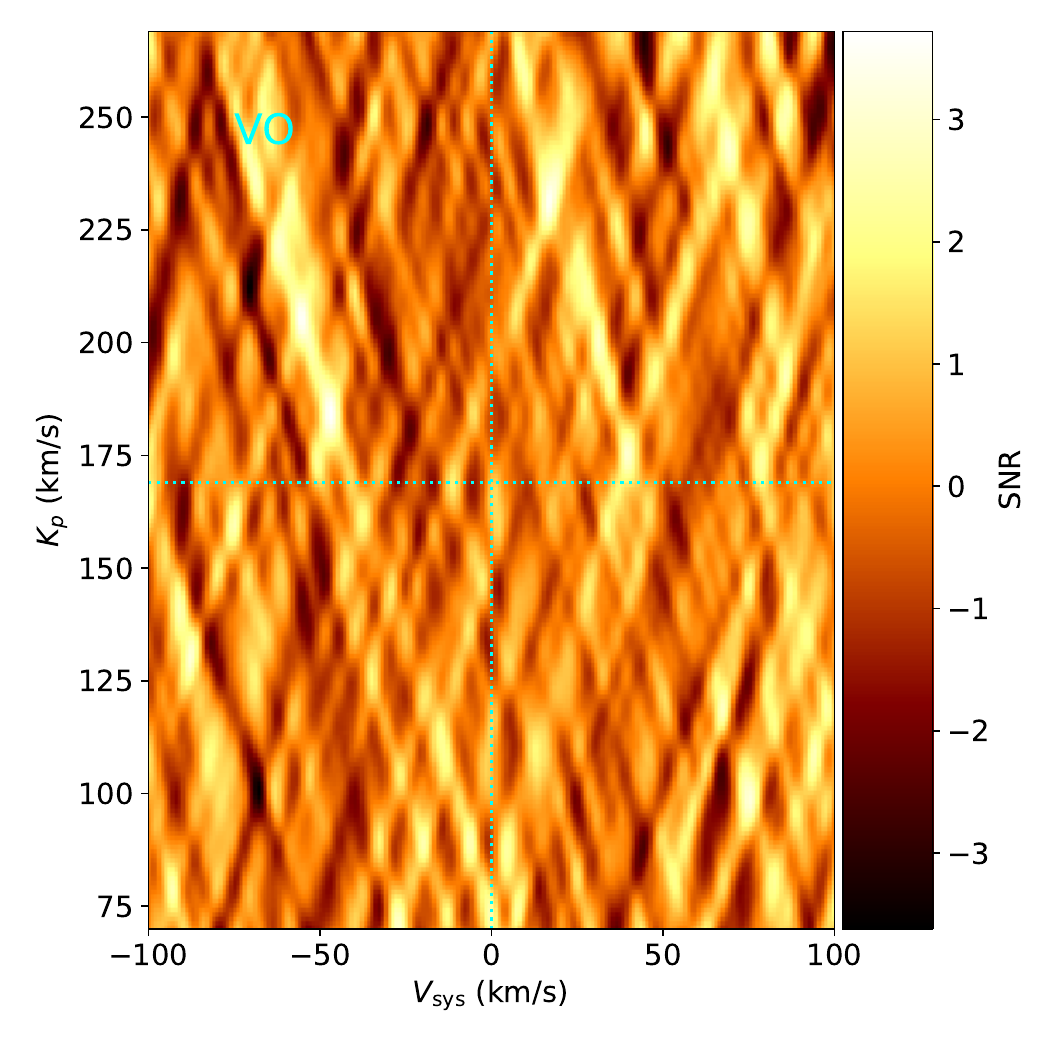}
\includegraphics[width=0.45\textwidth]{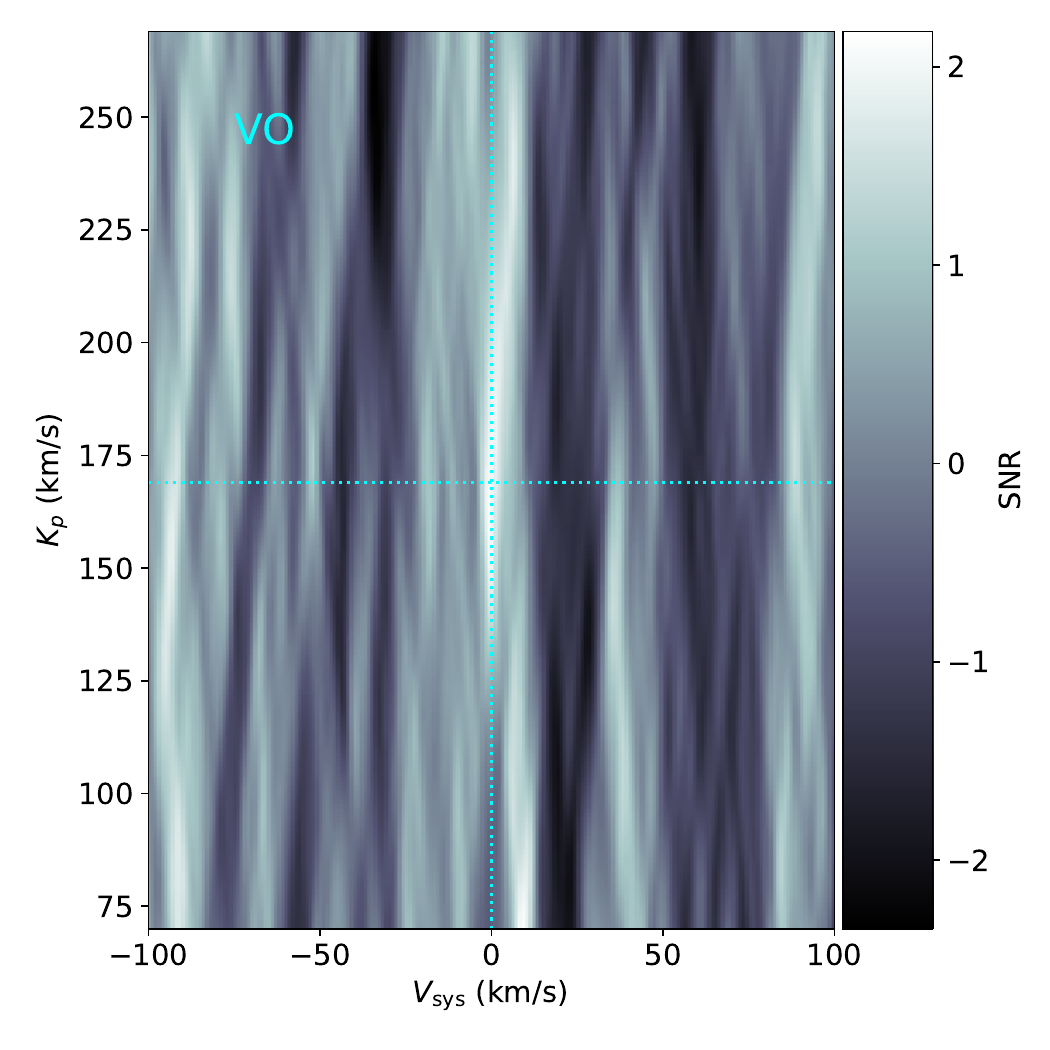}\\
\includegraphics[width=0.45\textwidth]{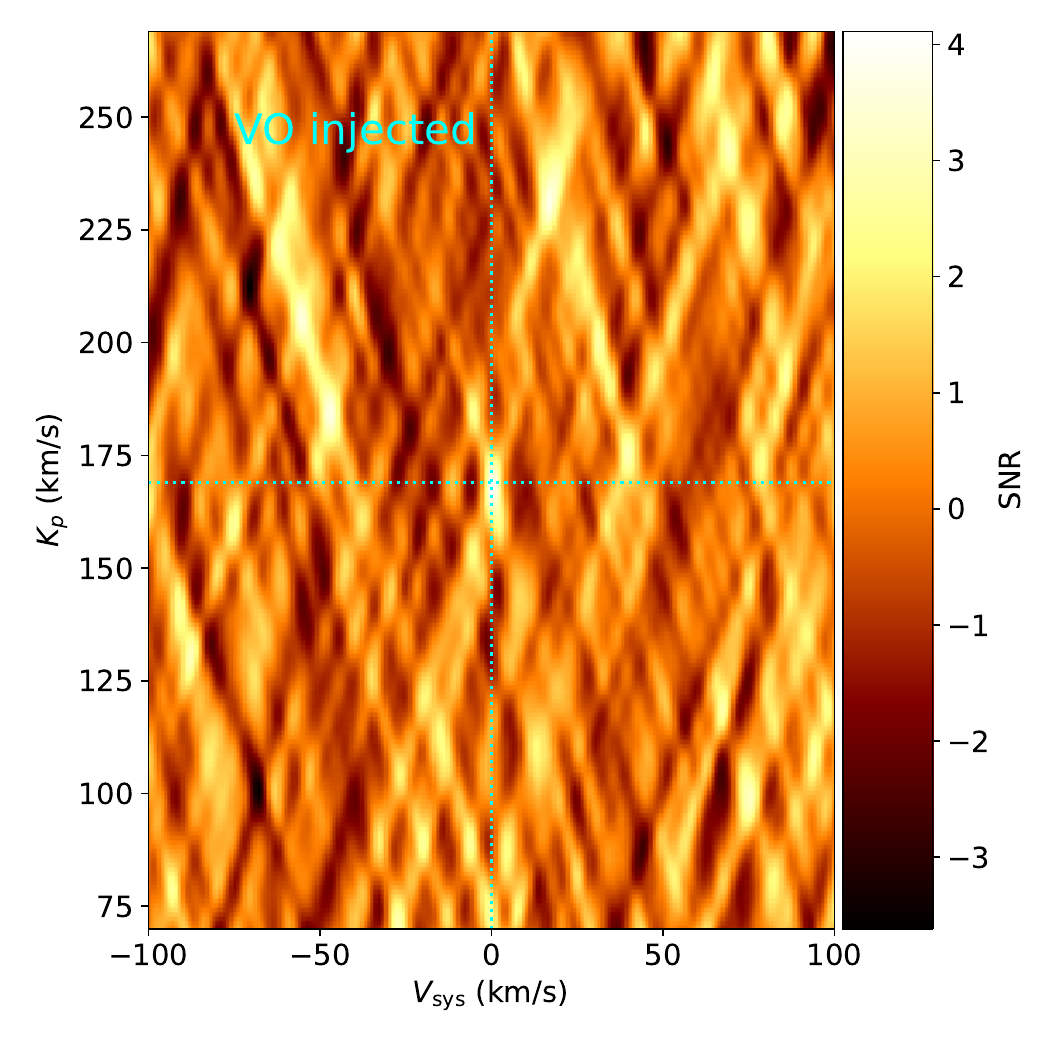}
\includegraphics[width=0.45\textwidth]{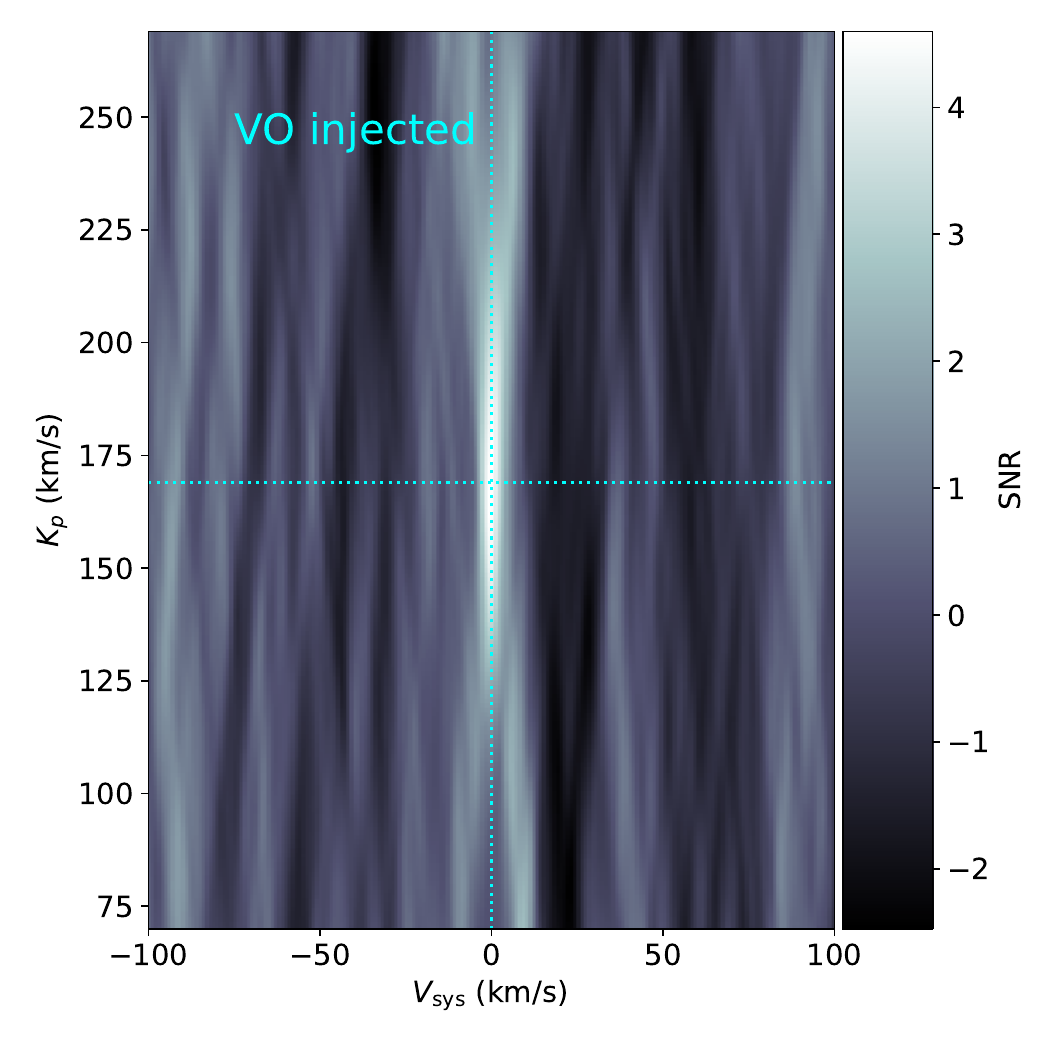}
\caption{Same as Fig.~\ref{fig:ccfs-tio} but for VO. We show the results for both injecting and recovering the ExoMol \cite{McKemmish16} line list.}
\label{fig:ccfs-vo}
\end{figure*}

\begin{figure*}
\centering
\includegraphics[width=0.45\textwidth]{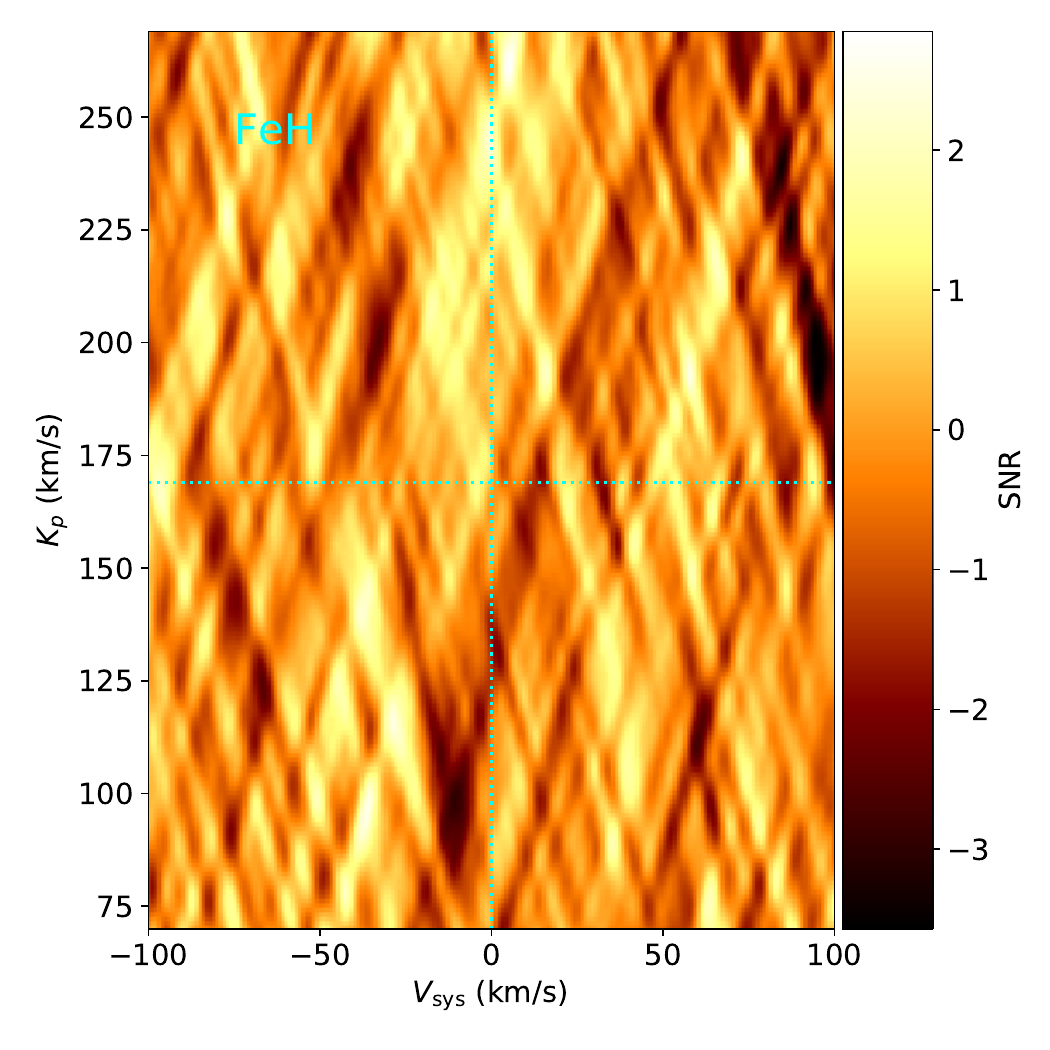}
\includegraphics[width=0.45\textwidth]{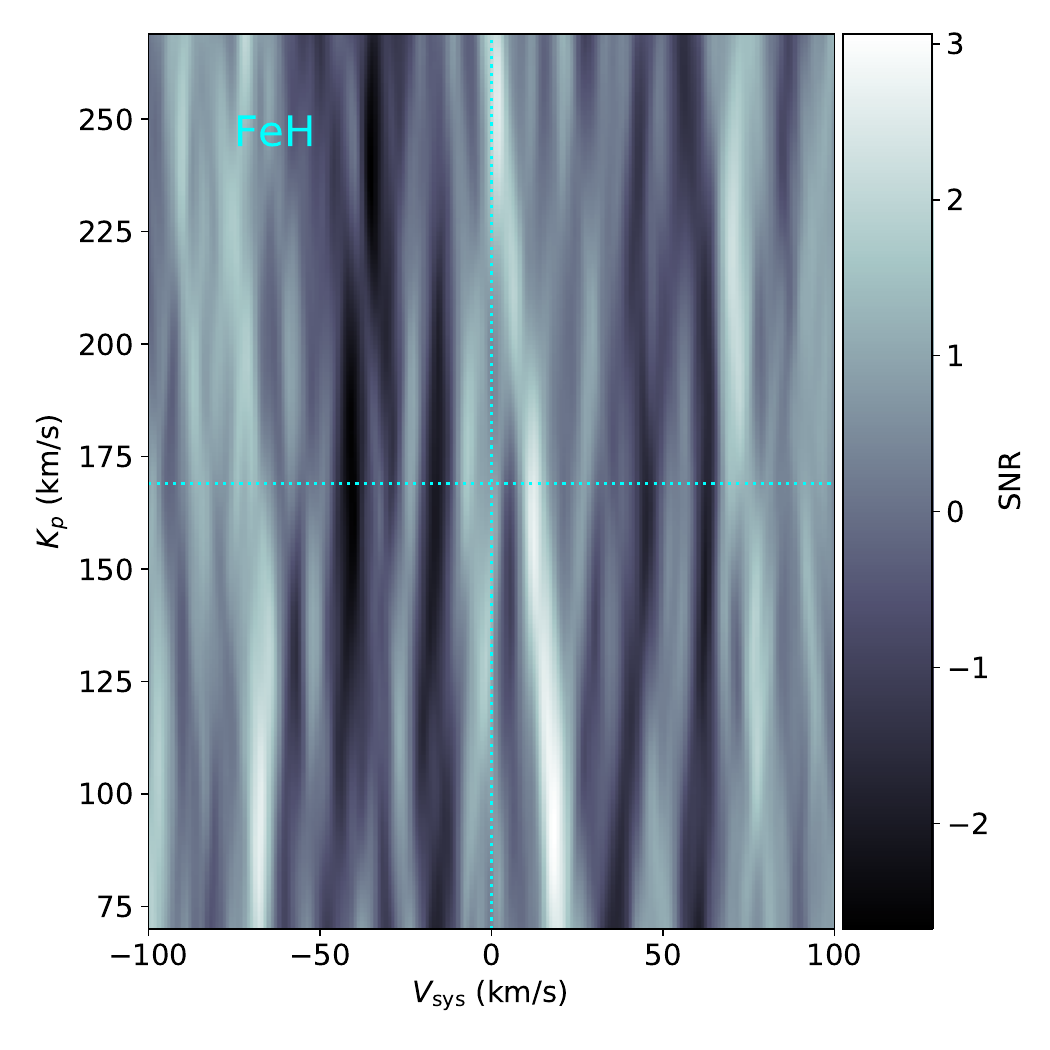}\\
\includegraphics[width=0.45\textwidth]{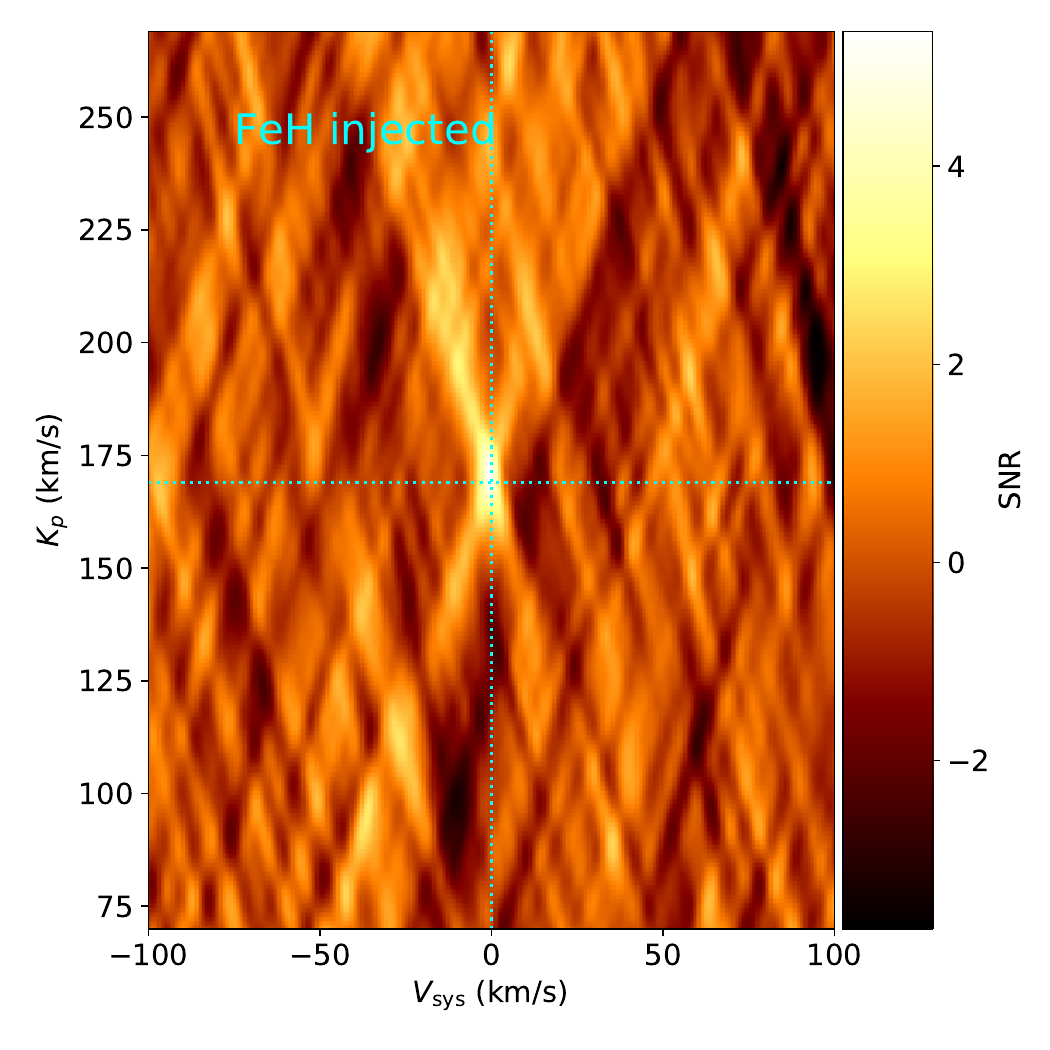}
\includegraphics[width=0.45\textwidth]{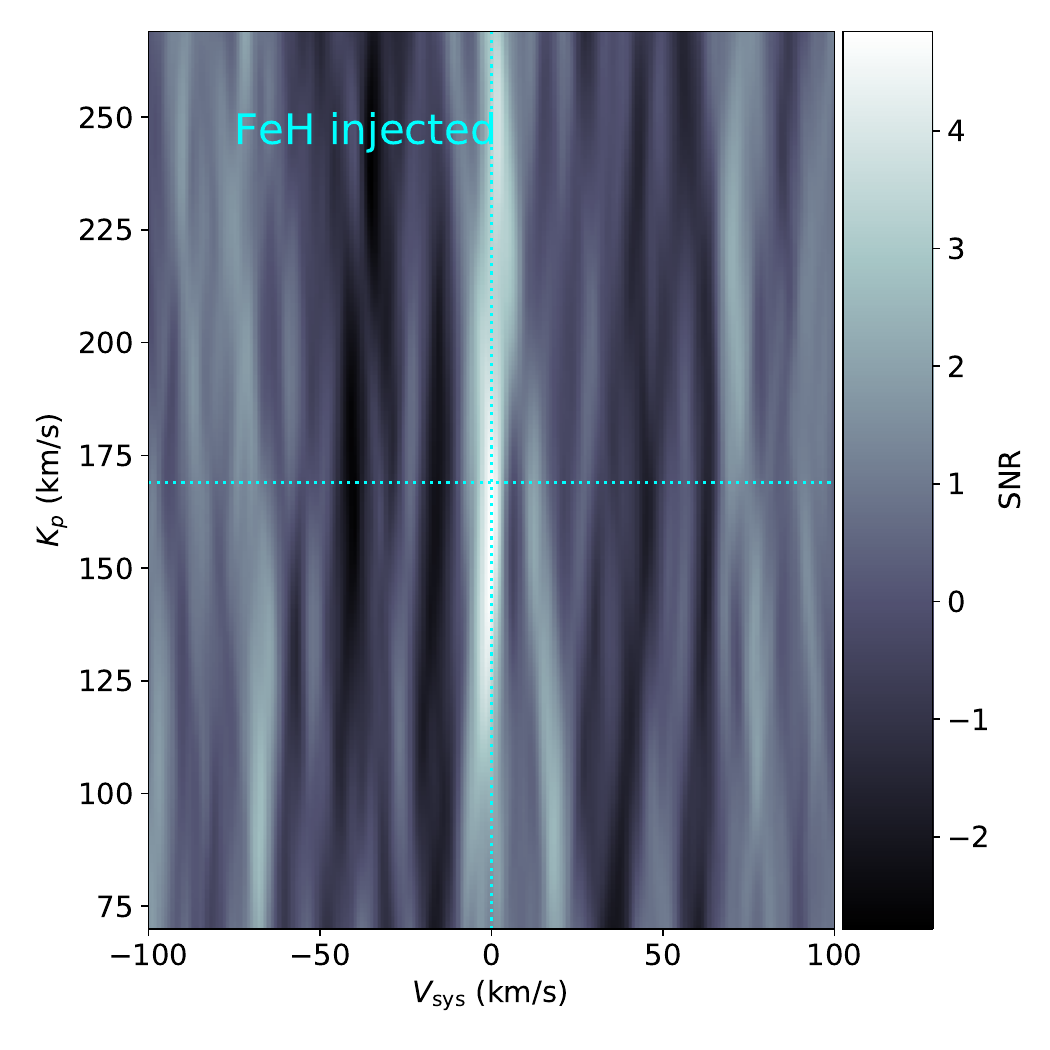}
\caption{Same as Fig.~\ref{fig:ccfs-tio} but for FeH.}
\label{fig:ccfs-feh}
\end{figure*}

\begin{figure*}
\centering
\includegraphics[width=0.45\textwidth]{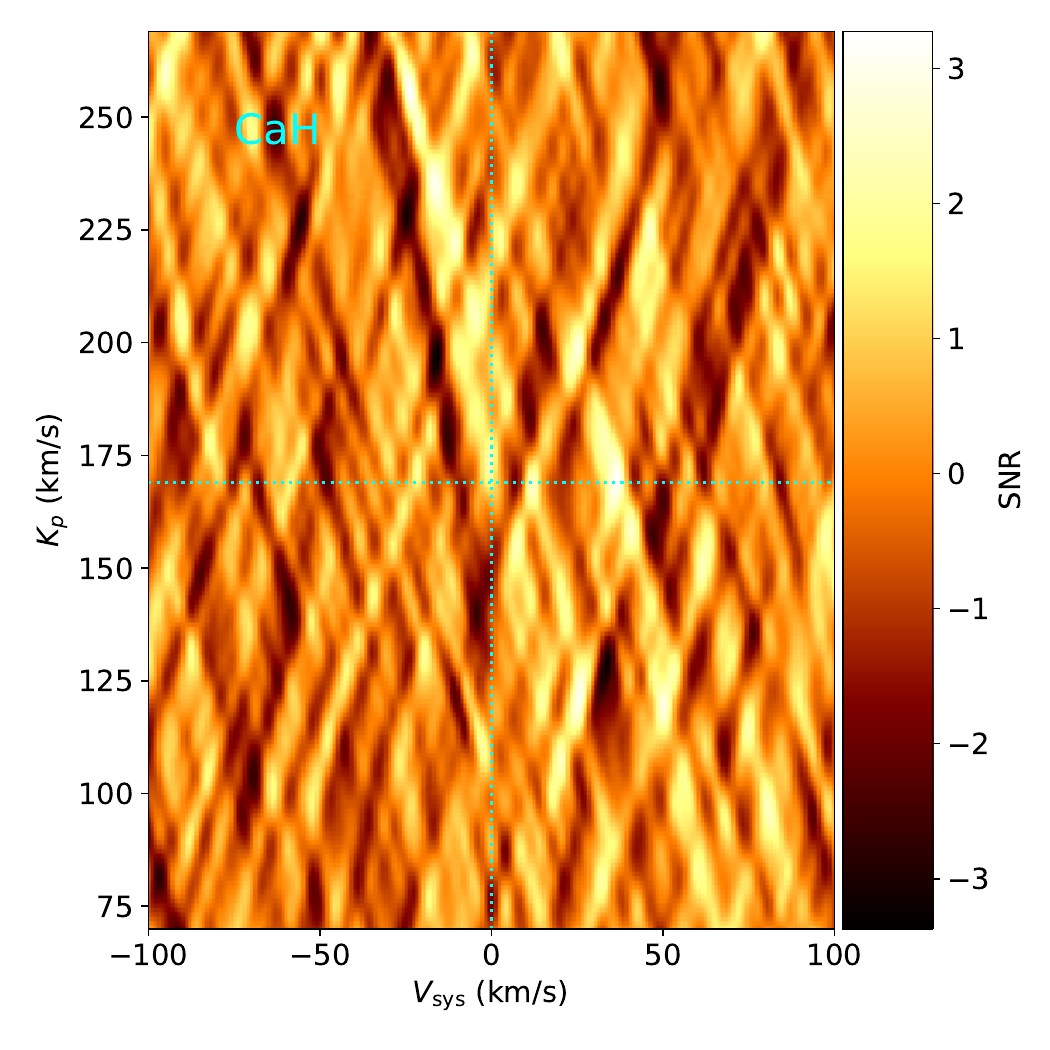}
\includegraphics[width=0.45\textwidth]{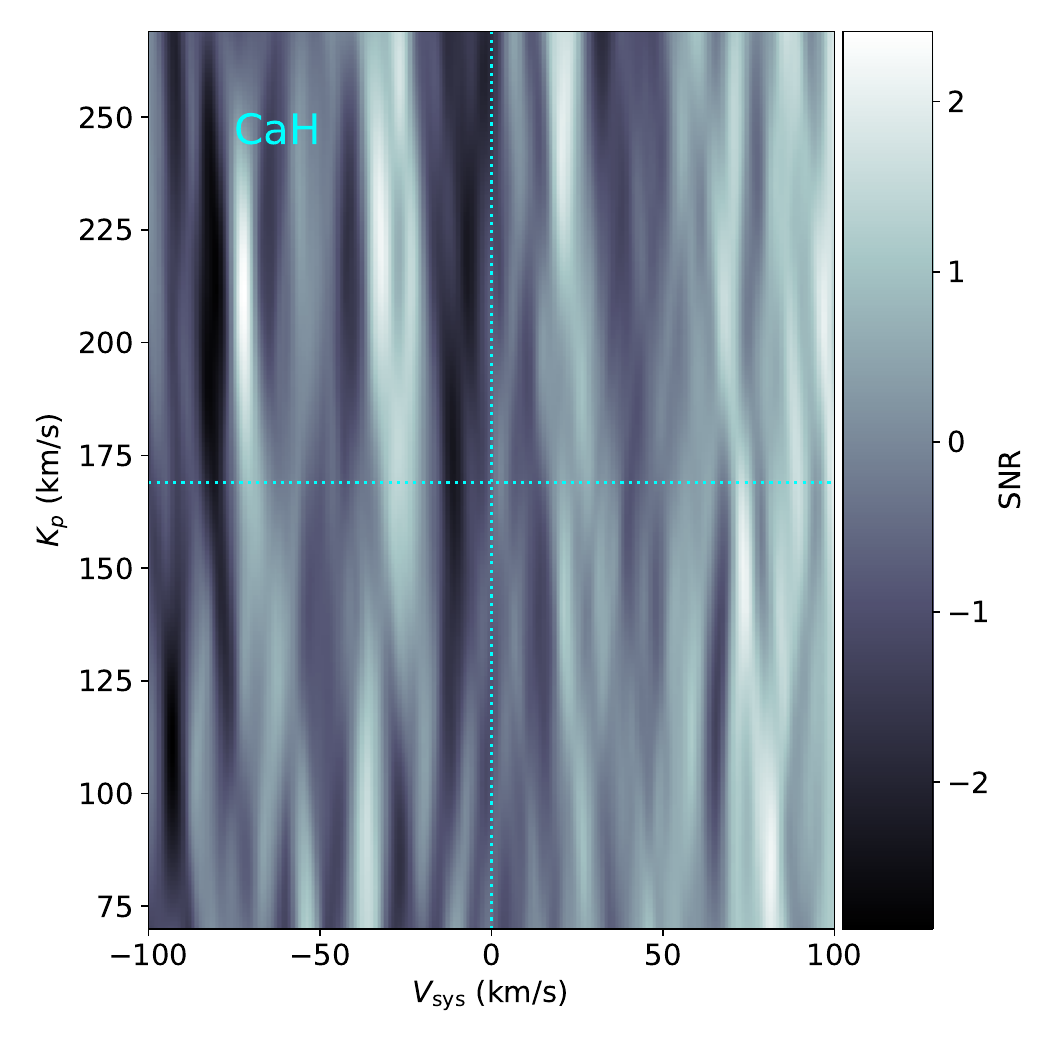}\\
\includegraphics[width=0.45\textwidth]{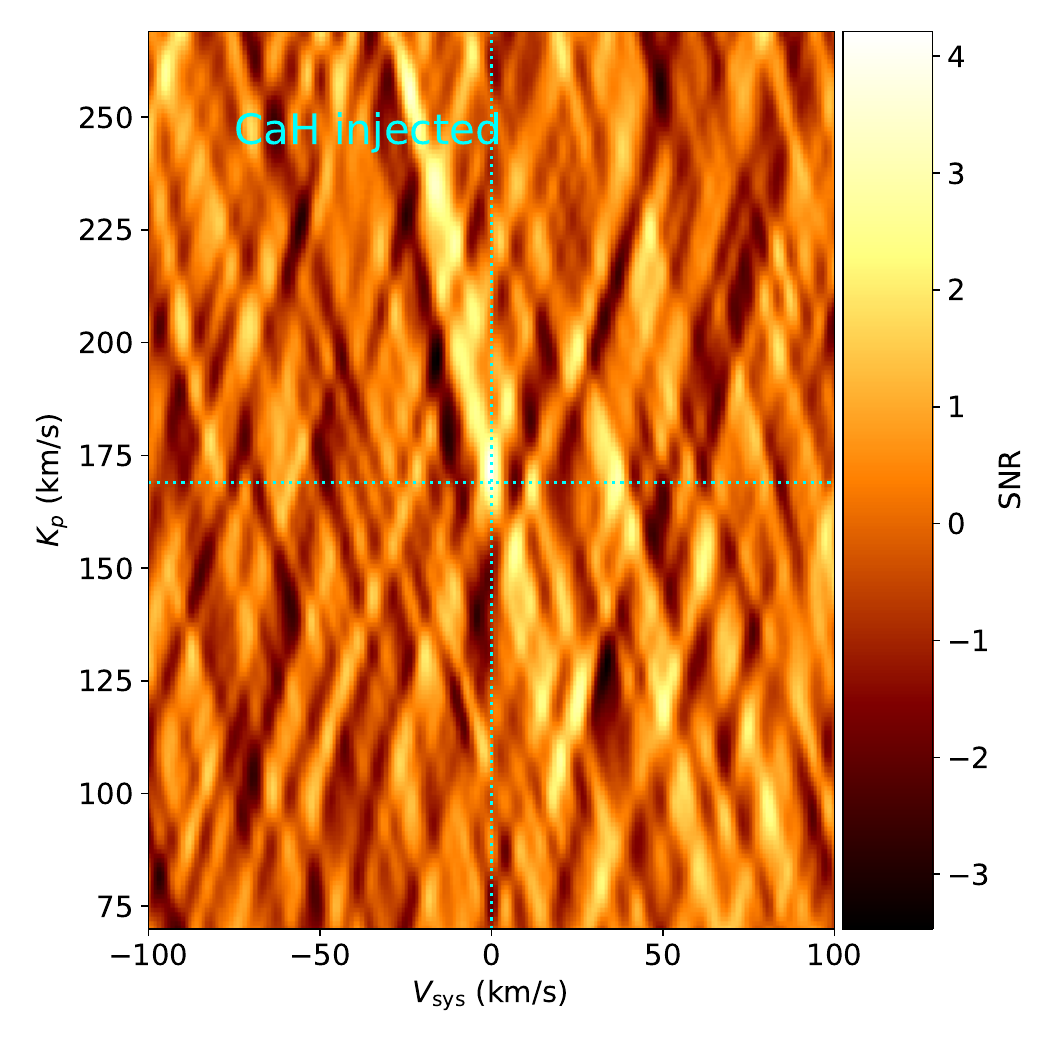}
\includegraphics[width=0.45\textwidth]{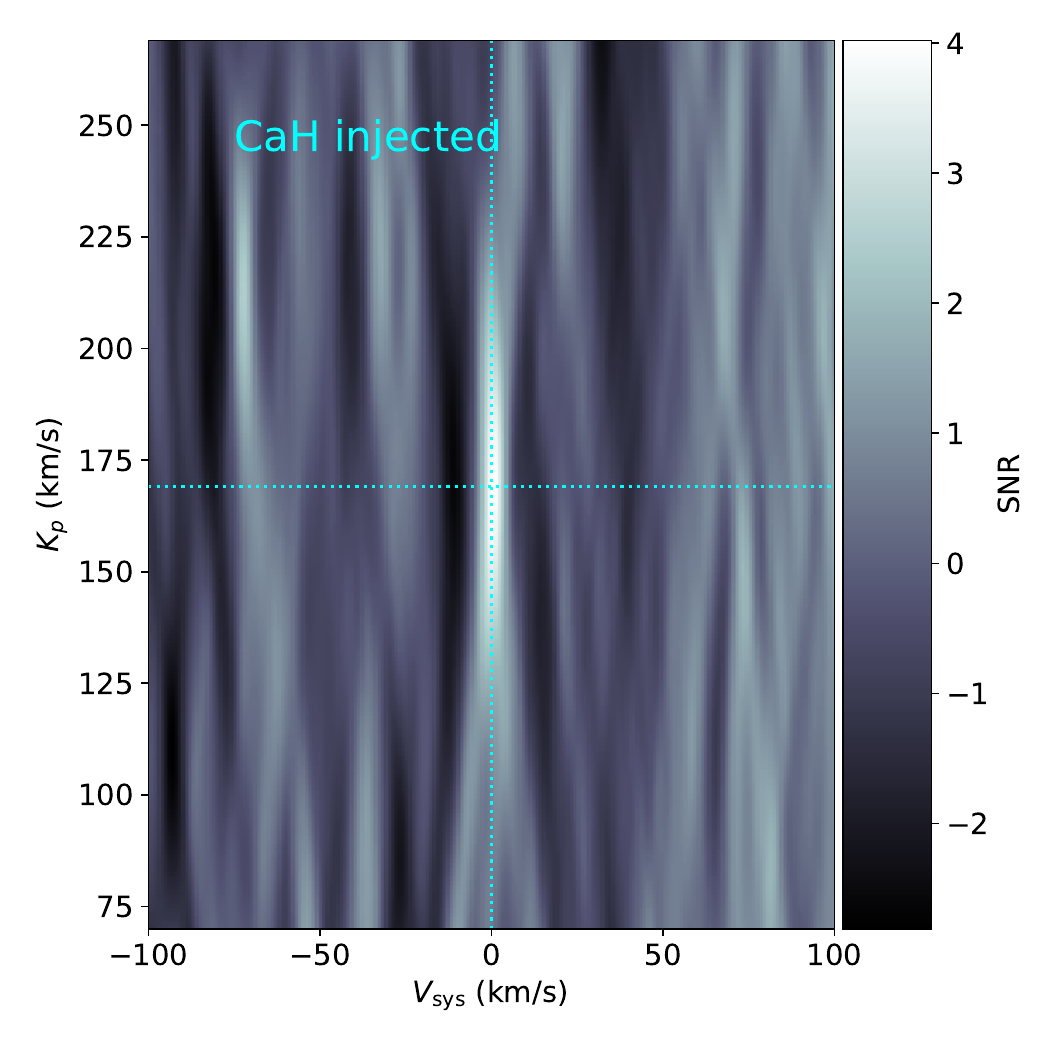}
\caption{Same as Fig.~\ref{fig:ccfs-tio} but for CaH.}
\label{fig:ccfs-cah}
\end{figure*}

In this Appendix we show the shifted and combined cross-correlation functions for the inversion agents TiO (Fig.~\ref{fig:ccfs-tio}), VO (Fig.~\ref{fig:ccfs-vo}), FeH (Fig.~\ref{fig:ccfs-feh}), and CaH (Fig.~\ref{fig:ccfs-cah}).

\end{CJK*}
\end{document}